\documentclass[sigconf,edbt]{acmart-edbt2025}

\usepackage{algorithmic}
\usepackage{multirow}
\usepackage{graphicx}
\usepackage{colortbl,hhline}
\usepackage{textcomp}
\usepackage{tikz}
\usepackage{xcolor}
\usepackage{adjustbox}
\usepackage{xspace}
\usepackage{enumitem}
\usepackage{subcaption}
\usepackage{pifont}
\usepackage{xurl}
\usepackage[]{mdframed}
\usepackage{framed}
\usepackage{listings}

\def\xmark{\textcolor{black}{\ding{53}}}
\def\cmark{\checkmark}
\def\dcmark{{\checkmark\kern-0.2em\checkmark}}

\def\bento{\textsf{Bento}\xspace}

\def\pandas{\textsf{Pandas}\xspace}

\def\polars{\textsf{Polars}\xspace}
\def\spark{\textsf{Spark}\xspace}
\def\koalas{\textsf{Koalas}\xspace}
\def\pyspark{\textsf{PySpark}\xspace}
\def\duckdb{\textsf{DuckDB}\xspace}
\def\sparksql{\textsf{SparkSQL}\xspace}
\def\sparkpd{\textsf{SparkPD}\xspace}
\def\dask{\textsf{Dask}\xspace}
\def\modin{\textsf{Modin}\xspace}
\def\modinr{\textsf{ModinR}\xspace}
\def\modind{\textsf{ModinD}\xspace}
\def\ray{\textsf{Ray}\xspace}
\def\vaex{\textsf{Vaex}\xspace}

\def\cudf{\textsf{CuDF}\xspace}
\def\datatable{\textsf{DataTable}\xspace}

\def\csv{\textsf{CSV}\xspace}
\def\parquet{\textsf{Parquet}\xspace}

\def\arrow{\textsf{Arrow}\xspace}
\def\docker{\textsf{Docker}\xspace}
\def\numpy{\textsf{NumPy}\xspace}
\def\python{\textsf{Python}\xspace}
\def\rust{\textsf{Rust}\xspace}

\def\scala{\textsf{Scala}\xspace}
\def\sql{\textsf{SQL}\xspace}
\def\athlete{\texttt{Athlete}\xspace}

\def\patrol{\texttt{Patrol}\xspace}

\def\taxi{\texttt{Taxi}\xspace}

\def\loan{\texttt{Loan}\xspace}

\def\BibTeX{{\rm B\kern-.05em{\sc i\kern-.025em b}\kern-.08em
    T\kern-.1667em\lower.7ex\hbox{E}\kern-.125emX}}

\usepackage{booktabs} % For formal tables

\setcopyright{rightsretained}

\acmDOI{https://doi.org/10.48786/edbt.2025.27}

\acmISBN{978-3-89318-098-1}

\acmConference[EDBT 2025]{28th International Conference on Extending Database Technology (EDBT)}{25th March-28th March, 2025}{Barcelona, Spain}
\acmYear{2025}

\begin{document}
\title{Evaluation of Dataframe Libraries for Data Preparation\\on a Single Machine}

\author{Angelo {Mozzillo}}
\affiliation{
  \institution{University of Modena\\and Reggio Emilia, Italy}
  \city{}
  \country{}
}
\email{angelo.mozzillo@unimore.it}

\author{Luca {Zecchini}}
\affiliation{
  \institution{University of Modena\\and Reggio Emilia, Italy}
  \city{}
  \country{}
}
\email{luca.zecchini@unimore.it}

\author{Luca {Gagliardelli}}
\affiliation{
  \institution{University of Modena\\and Reggio Emilia, Italy}
  \city{}
  \country{}
}
\email{luca.gagliardelli@unimore.it}

\author{Adeel {Aslam}}
\affiliation{
  \institution{University of Modena\\and Reggio Emilia, Italy}
  \city{}
  \country{}
}
\email{adeel.aslam@unimore.it}

\author{Sonia {Bergamaschi}}
\affiliation{
  \institution{University of Modena\\and Reggio Emilia, Italy}
  \city{}
  \country{}
}
\email{sonia.bergamaschi@unimore.it}

\author{Giovanni {Simonini}}
\affiliation{
  \institution{University of Modena\\and Reggio Emilia, Italy}
  \city{}
  \country{}
}
\email{giovanni.simonini@unimore.it}

\renewcommand{\shortauthors}{Angelo Mozzillo et al.}

\begin{abstract}
\emph{Data preparation} is a trial-and-error process that typically involves countless iterations over the data to define the best pipeline of operators for a given task.
With tabular data, practitioners often perform that burdensome activity on local machines by writing ad hoc scripts with libraries based on the \pandas dataframe API and testing them on samples of the entire dataset---the faster the library, the less idle time its users have.

In this paper, we evaluate the most popular \python dataframe libraries in general data preparation use cases to assess how they perform on a single machine.
To do so, we employ 4 real-world datasets with heterogeneous features, covering a variety of scenarios, and the TPC-H benchmark.
The insights gained with this experimentation are useful to data scientists who need to choose which of the dataframe libraries best suits their data preparation task at hand.

In a nutshell, we found that:
for small datasets, \pandas consistently proves to be the best choice with the richest API;
when data fits in RAM and there is no need for complete compatibility with \pandas API, \polars is the go-to choice thanks to its in-memory execution and query optimizations;
when a GPU is available, \cudf often yields the best performance, while for very large datasets that cannot fit in the GPU memory and RAM, \pyspark (thanks to a multithread execution and a query optimizer) proves to be the best option.
\end{abstract}

\maketitle

\section{Introduction}
\label{sec:introduction}

Companies and organizations significantly depend on their data to drive informed decisions, such as business strategy definition, supply chain management, etc.
Thus, guaranteeing high data quality is fundamental to ensure the reliability of analysis and avoid undesired additional costs~\cite{haug2011data_quality@jiem}.
Data is often heterogeneous, i.e., data sources adopt different formats and conventions (e.g., different encodings, different ways to represent dates or numeric values, etc.), and are affected by quality issues, such as missing or duplicate values~\cite{fan2015data_quality@sigmod_record}.
Due to that, data scientists dedicate considerable time and resources to perform \emph{data preparation}~\cite{fernandes2023data_preparation@sncs}.
Typically, this fundamental process involves multiple operations called \emph{preparators}~\cite{hameed2020data_preparation_survey@sigmod_record}, combined into a pipeline~\cite{furche2016data_wrangling@edbt}, aimed at exploring~\cite{lissandrini2018data_exploration@book}, cleaning~\cite{ilyas2019data_cleaning@book}, and transforming raw data into curated datasets~\cite{terrizzano2015data_wrangling@cidr, rezig2019data_civilizer_demo@pvldb, geerts2020llunatic@vldbj}.

It is said that data scientists spend up to 80\% of their time on data preparation~\cite{hellerstein2018data_preparation@debull}.
This also depends on the lack of support in the choice among many different available libraries and in the design of the pipeline that best suits both the dataset and the task at hand.
The growing need for standardization and best practices has driven an increasing number of data practitioners to embrace the \emph{dataframe} as the foundational data structure for constructing pipelines and developing libraries.
At a high level, the dataframe is a two-dimensional data structure that consists of rows and columns~\cite{petersohn2020modin@pvldb}.
Libraries for data preparation typically offer a flexible functional interface that allows to conveniently compose pipelines.

\pandas~\cite{mckinney2010pandas@scipy} is by far the most widely adopted library for manipulating dataframes, and is considered by many as the \emph{de facto} standard for all preparators~\cite{petersohn2020modin@pvldb, jindal2021magpie, psallidas2022data, leelux}.
This library has become so popular in the data science ecosystem that even the popularity of \python itself has been sometimes attributed to its wide adoption~\cite{claburn2017python_explosion@the_register}.

Despite its success, \pandas notoriously presents multiple severe limits~\cite{DBLP:conf/edbt/EmaniFC24, DBLP:conf/btw/KlabeH21}.
Firstly, it has not been designed to work with large datasets efficiently.
In fact, \pandas operates in a single-threaded manner, lacks support for cluster deployment, and does not implement any memory optimization mechanism (the entire data is kept in main memory until the end of the pipeline execution).
To overcome these limitations, several alternative libraries have been developed and are gaining popularity among data scientists.
However, these libraries come with heterogeneous features (e.g., lazy evaluation, GPU support, etc.), making it hard for data scientists to navigate among them and choose the most suitable solution.

% {\color{red} A major reason for this confusion} is the lack of extensive and rigorous studies to evaluate the performance of such libraries in the data preparation scenario.
In particular, data scientists lack the support of extensive and rigorous studies to evaluate the performance of such libraries in the data preparation scenario.
Public wisdom about dataframe libraries is mostly scattered across several not peer-reviewed sources doing mutual comparisons~\cite{schmitt2020scaling_pandas@data_revenue, alexander2023beyond_pandas@medium, pinner2023dataframes@medium} (whose major claims are mostly confirmed by our evaluation), while related examples in scientific literature~\cite{petersohn2021modin@pvldb, rehman2022fuzzydata@testdb, shanbhag2023dataframe_energy_consumption@msr} either focus on orthogonal aspects or only cover small subsets of libraries and preparators.
% polars_alternatives@polars
% Similarly, relational benchmarking systems such as TPC-H\footnote{\url{https://www.pola.rs/benchmarks.html}} offer limited insights about data preparation workflows~\cite{rehman2022fuzzydata@testdb}.

\subsubsection*{Our contribution}
It is common for data scientists to perform countless iterations on samples of the datasets on their laptop or PC to define a proper data preparation pipeline for the task at hand.
Hence, the faster the employed library is, the less idle time they have.
To support data scientists in the choice of the library that can speedup their workflow, we present the first extensive experimental comparison of \pandas and its most popular dataframe-based alternatives, namely \pyspark, \modin, \polars, \cudf, \vaex, and \datatable.
For the sake of reproducibility we developed \bento, a framework that provides a general interface to design and deploy the data preparation pipeline with any library.
\bento\footnote{\url{https://github.com/dbmodena/bento}} allows the definition of a common \pandas-like API for all libraries and enables the execution in a containerized environment using Docker.
This process is streamlined with the help of a convenient configuration file\footnote{\url{https://github.com/dbmodena/bento/tree/master\#write-a-test-file}}.

To cover a variety of scenarios and to provide results that are representative of real-world use cases, we consider four datasets with heterogeneous size and features, previously adopted in related work on data preparation~\cite{hameed2020data_preparation_survey@sigmod_record} and dataframe libraries~\cite{petersohn2021modin@pvldb}.
All these datasets are collected from Kaggle~\cite{kaggle}, the most popular platform for data science challenges.
This allows us to exploit published data preparation pipelines that have proven to work well for downstream machine learning tasks and have been validated by the world's largest data science community.
% \footnote{\url{https://www.kaggle.com}}
Moreover, we further validate our findings by testing the different libraries with the TPC-H benchmark~\cite{poss2000tpc_benchmark@sigmod_record}.

We identify four essential data preparation stages: input/output (I/O), exploratory data analysis (EDA), data transformation (DT), and data cleaning (DC).
Thus, we evaluate the performance of the libraries both on a single preparator and on sequences of preparators (i.e., the entire pipeline or its subsets corresponding to the four different stages), highlighting the impact of the query optimization techniques supported by some of them.

% \begin{figure}[!t]
% \centering
%      \includegraphics[width=\columnwidth]{input/images/Picture 1.png}
%      % \vspace*{-0.75cm}
%      \caption{Overview of the Bento framework.}
%      \label{fig:bento}
% \end{figure}

\subsubsection*{For whom is this paper relevant}
Since it represents by far the most common situation for data scientists~\cite{DBLP:conf/edbt/EmaniFC24}, we only focus on the single-machine scenario, aiming to explore the distributed scenario in a future extension\footnote{Note that we decided to include libraries conceived for distributed execution, such as \pyspark or \modin, since through multithreading and optimized execution plans they can improve the performance of dataframe operations even on a single machine.}.
Our findings are useful to those data scientists that: \emph{(i)} are employing \pandas in their workflow when devising data preparation pipelines on their laptop or PC; \emph{(ii)} want to replace \pandas with a more efficient library that has the same interface.
Thus, we evaluate scalability by measuring the performance of the dataframe libraries on three distinct machine configurations (laptop, workstation, and server), while increasing the size of the dataset.
Hence, we provide the readers with several insights about their performance based on multiple factors, such as the size and the features of the dataset, the configuration of the underlying machine, and the type of preparators applied.

\subsubsection*{Outline}
We describe in detail the dataframe data structure and the libraries to be compared in Section~\ref{sec:preliminaries}, then we present our evaluation setup and the results in Section~\ref{sec:dsandpipe} and~\ref{sec:results}, respectively.
After reviewing the related literature in Section~\ref{sec:related_work}, we report the key takeaways in Section~\ref{sec:key_takeaways} and draw the conclusions in Section~\ref{sec:conclusion}.

% -------------------------------------------------- %

\section{Dataframes}
\label{sec:preliminaries}

A \emph{dataframe}~\cite{petersohn2020modin@pvldb} is a data structure that organizes the data into a two-dimensional table-like format, where columns represent the schema and rows represent the content.
Each column has a specific data type, and the schema does not need to be declared in advance.
A dataframe provides a wide set of operators acting on all data dimensions, enabling and simplifying various data preparation tasks such as data analysis, transformation, and cleaning.
Beyond this variety of features, dataframes became popular thanks to their simple and flexible structure, which allows them to effectively handle different types of data.
In fact, compared to specific \emph{data preparation tools}~\cite{hameed2020data_preparation_survey@sigmod_record}, dataframes offer greater simplicity and customization, enabling users to manipulate the data by writing \python code without the effort of learning to use a new tool.

% -------------------------------------------------- %

\pandas~\cite{pandas_github} has been the first widely-adopted \python library implementing dataframes (inspired by the corresponding data structure from the \textsf{R} language~\cite{mckinney2010pandas@scipy}) and provides efficient and expressive data structures optimized to work with structured datasets.
% \footnote{\url{https://github.com/pandas-dev/pandas}}
\pandas has established itself over time as the standard library for data manipulation and analysis.
The main reasons behind its widespread popularity reside in its robust functionality and user-friendly design, making it the go-to choice for handling datasets in \python.
Given its enormous diffusion, \pandas is a very mature library and the existence of a huge amount of documentation and courses makes its learning extremely straightforward.

Nevertheless, \pandas also presents several notorious limitations~\cite{mckinney2017ten_things_i_hate@blog}, which become especially evident when dealing with large-scale datasets, as it lacks the optimizations required to process large amounts of data in an efficient way.
For instance, \pandas does not support multiprocessing and parallel computing, and it is designed to work with in-memory datasets, lacking support for datasets that exceed memory limits.
Further, it needs to materialize the intermediate result of every operation (this execution strategy is known as \emph{eager evaluation}), exposing it to out-of-memory risks and preventing it from applying query optimization techniques.

% While \pandas initially relied on \numpy\footnote{\url{https://github.com/numpy/numpy}} data structures for memory management, it moved to \textsf{Apache Arrow}\footnote{\url{https://github.com/apache/arrow}} starting from \pandastwo\footnote{\url{https://pandas.pydata.org/docs/dev/whatsnew/v2.0.0.html}}.
% \arrow is a language-independent columnar memory format for flat and hierarchical data, organized to optimize analytic operations on modern hardware, allowing \pandastwo to handle large datasets more efficiently.

To address such limitations, many solutions have emerged to seamlessly replace \pandas in \python workflows, aiming for more efficient data processing.
% In the following, we describe several \python libraries alternative to \pandas.
The selection of the \python libraries to include in our evaluation was based on three main criteria, requiring them to:
\emph{(i)} be based on the dataframe data structure (excluding therefore \sql-based solutions such as \textsf{DuckDB}~\cite{raasveldt2019duckdb_demo@sigmod});
\emph{(ii)} be compatible with the Pandas API, to enable a fair and straightforward evaluation of each operation (note that many of these libraries, recognizing \pandas as the \emph{de facto} standard, already aim at Pandas-compatibility for their API~\cite{modin_documentation} or describe the parallelism between their API and the Pandas API~\cite{datatable_pandas_comparison});
% \footnote{\url{https://modin.readthedocs.io/en/latest}}
% \footnote{\url{https://datatable.readthedocs.io/en/latest/manual/comparison_with_pandas.html}}
\emph{(iii)} have gained more than 1k stars on GitHub, denoting the existence of a significant user base and community support.
The selected libraries are: \pandas, \pyspark, \modin, \polars, \cudf, \vaex, and \datatable.
% We compare them highlighting their strengths and weaknesses, to facilitate efficient data processing and provide support to data scientists and practitioners for efficiently handling their data preparation workflows.

% \usepackage{graphicx}
\begin{table*}[ht]
\small
\caption{Features of the compared dataframe libraries.}
\label{tab:libraries}
% \resizebox{\textwidth}{!}{%
\begin{tabular}{r|c|c|c|c|c|c|c|}
\cline{2-8}
\multicolumn{1}{c|}{} &
  \textbf{Pandas} &
  \textbf{PySpark} &
  \textbf{Modin} &
  \textbf{Polars} &
  \textbf{CuDF} &
  \textbf{Vaex} &
  \textbf{DataTable} \\ \hhline{-=======}
\multicolumn{1}{|r|}{Multithreading}     &             & \checkmark   & \checkmark & \checkmark &            & \checkmark & \checkmark \\ \hline
\multicolumn{1}{|r|}{GPU acceleration}   &             &              &            &            & \checkmark &            &            \\ \hline
\multicolumn{1}{|r|}{Resource optimization} &
   &
  \checkmark &
  \checkmark &
  \checkmark &
  \checkmark &
  \checkmark &
  \checkmark \\ \hline
\multicolumn{1}{|r|}{Lazy evaluation}    &             & \checkmark   &            & \checkmark &            &            &            \\ \hline
\multicolumn{1}{|r|}{Deploy on cluster}  &             & \checkmark   & \checkmark &            &            &            &            \\ \hline
\multicolumn{1}{|r|}{Native language} &
  Python &
  Scala &
  Python &
  Rust &
  C\char`/ C++ &
  C\char`/ Python &
  C++\char`/ Python \\ \hline
\multicolumn{1}{|r|}{Licence}          & 3-Clause BSD & Apache 2.0 & Apache 2.0           &   MIT  &  Apache 2.0  & MIT              &   Mozilla Public 2.0         \\ \hline
\multicolumn{1}{|r|}{Other requirements} &             & SparkContext & Ray\char`/ Dask   &            & CUDA       &            &            \\ \hline
\multicolumn{1}{|r|}{Considered version} & 2.2.1 & 3.5.1        & 0.29.0     & 0.20.23     & 24.04.01    & 4.17.0     & 1.1.0      \\ \hline
\end{tabular}%
%}
\end{table*}

\subsection*{Dataframe Libraries}

We identify five major features exploited by \pandas alternatives to achieve high efficiency:

\begin{itemize}[]
    \item \emph{Multithreading}: using multithreading to speed up the execution of dataframe operations, making the most of the available hardware resources.
    \item
    \emph{GPU acceleration}: leveraging the parallel computing potential of graphics processing units to further improve the performance of dataframe operations.
    \item
    \emph{Resource optimization}: implementing strategies to efficiently manage memory, reducing the impact on the available resources, hence improving the overall performance.
    \item
    \emph{Lazy evaluation}: maintaining a logical plan of the operations, triggering their execution only when a specific output operation is invoked, to apply query optimization techniques.
    \item
    \emph{Deploy on cluster}: enabling the distribution of dataframe processing across a cluster of machines, leveraging parallel computing (as stated in Section~\ref{sec:introduction}, we aim to consider this dimension in future work).
\end{itemize}

For each dataframe library, Table \ref{tab:libraries} lists which of these features it implements, along with information about its implementation and the version considered here.
% Table \ref{tab:libraries} lists for each dataframe library which of these features it implements, along with other characteristics about its implementation and the version considered here.

% -------------------------------------------------- %

\smallskip
\noindent
\textbf{\pyspark}~\cite{pyspark_github} is the \python API for \textsf{Apache Spark}, which is primarily implemented in \scala.
% \footnote{\url{https://github.com/apache/spark}}

% \spark allows to work with distributed collections of data called \emph{Dataset}, which can be built from \java objects and manipulated using functional transformations.
\pyspark \textsf{DataFrame} is a type of \textsf{Dataset} (i.e., a distributed collection of data) organized into named columns, resembling a relational table and supporting relational operators~\cite{pyspark_dataframes}.
% \footnote{\url{https://spark.apache.org/docs/latest/sql-programming-guide.html}}
\pyspark also supports lazy evaluation, relying on the \textsf{Catalyst} optimizer~\cite{armbrust2015spark_sql@sigmod} and a disk spillover mechanism that automatically offloads data from RAM to disk when memory limits are reached~\cite{pyspark_diskspill}. This allows \pyspark to process datasets that exceed the machine's physical memory capacity.
% that translates the applied pipeline of operators into a query tree of logical operators, optimized to generate the final execution plan.
While \spark owes its popularity to the capability of processing large-scale datasets on a cluster of machines, \emph{standalone} (i.e., single-machine) mode is not only supported, but also surprisingly fast in multiple scenarios, as shown in Section~\ref{sec:results}.
% (represented within \spark as a \textrm{SparkContext} object)

\pyspark supports two different APIs:
\emph{(i)} \textsf{Spark SQL}, which allows combining \sql queries with \spark programs to work with structured data;
% , to efficiently read, write, transform, and analyze data using \python and \sql;
\emph{(ii)} \textsf{Pandas on Spark} (denoted as \textsf{SparkPD} and previously known as \textsf{Koalas}~\cite{koalas_github}), which enables (by adding an index to the conventional \spark dataframe) to distribute \pandas workloads across multiple nodes without requiring modifications to the original \pandas code for most API functions ($\sim$80\%).
% before its inclusion in \pyspark starting from \spark 3.2
% \footnote{\url{https://github.com/databricks/koalas}}

% -------------------------------------------------- %

\smallskip
\noindent
\textbf{\modin}~\cite{modin_github} is a \python library that provides a parallel alternative to \pandas~\cite{petersohn2020modin@pvldb, petersohn2021modin@pvldb}.
% \footnote{\url{https://github.com/modin-project/modin}}
\modin adopts the \pandas data format as the default storage layer and employs a set of 15 \textit{core operators} to simplify \pandas functions and build its own \pandas-like API.
When core operators cannot handle a function, it switches to the \emph{default to \pandas} mode (affecting performance due to communication costs and \pandas single-threaded nature), reverting to a partitioned \modin dataframe after completion.
It also implements \emph{opportunistic evaluation}~\cite{xin2021opportunistic_evaluation@debull}, enabling execution based on interactions, and facilitates incremental query construction through intermediate dataframe results.
% , prioritizing the materialization of small results and employing predictive mechanisms.
\modin is designed to dynamically switch between different partition schemes (row-based, column-based, or block-based) depending on the operation.
Each partition is then processed independently by the execution engine: \dask or \ray (we consider both solutions, denoted as \modind and \modinr, respectively).
% \modin adopts the \pandas data format as the default storage layer, ensuring compatibility and usability, and parallelizes the execution of the operators across multiple cores~\cite{petersohn2021modin@pvldb}.

% \modin employs a set of 15 main operators, called \textit{core operators}, to simplify \pandas functions and build its own \pandas-like API. 
% When core operators cannot handle a function, \modin switches to the \emph{default to \pandas} mode (incurring performance costs due to communication and \pandas single-threaded nature), then it reverts to a partitioned \modin dataframe after completion.
% \modin addresses flexible dataframe schemas by revising schema induction rules, allowing schema omission when not needed, optimizing queries, and minimizing overhead. 
% The system implements \emph{opportunistic evaluation}~\cite{xin2021opportunistic_evaluation@debull}, enabling execution based on interactions. 
% Incremental query construction is facilitated through intermediate dataframe results, prioritizing materialization of small results and employing predictive mechanisms. 
% \modin captures and stores ordering information separately for on-demand ordering, and it explores key-value pairs for the efficient storage of sparse dataframes.

% Since \modin can run using two different engines, \dask and \ray, we consider both in the evaluation (denoted as \modind and \modinr, respectively).

\dask~\cite{dask_github} is a \python library for distributed computing, which allows working with large distributed \pandas dataframes~\cite{rocklin2015dask@scipy}.
% \footnote{\url{https://github.com/dask/dask}}
It is designed to efficiently extend memory capacity using disk space, making it suitable for systems with limited memory, and its scheduler provides flexibility to the execution.
% \dask uses dictionaries, tuples, and callables to represent task schedules.
% The scheduler can operate in different modes, providing flexibility to the execution.
While we also considered the inclusion of \dask as an independent library, we do not report its results since we found it to perform better in combination with \modin (as \dask is not well suitable for a single machine) and it only covers $\sim$55\% of the \pandas API (while \modin covers $\sim$90\%)~\cite{modin_and_dask}.
% We considered the inclusion of \dask in our evaluation also as an independent library, but we found it to perform better in combination with \modin (\dask is not well-suited for a single machine), hence we do not report its results here.
% Furthermore, \dask covers approximately 55\% of the \pandas API, while \modin around 90\%.

\ray~\cite{ray_github} is a general-purpose framework for parallelizing \python code, using an in-memory distributed storage system and \textsf{Apache Arrow}~\cite{arrow_github} (a language-independent columnar memory format, recently adopted even by \pandas~\cite{pandas2_news}) as data format~\cite{moritz2018ray@osdi}.
% \footnote{\url{https://github.com/ray-project/ray}}
% \footnote{\url{https://github.com/apache/arrow}}
% \footnote{\url{https://pandas.pydata.org/docs/dev/whatsnew/v2.0.0.html}}
Unlike \dask, \ray does not have built-in primitives for partitioned data.
While the two engines serve different use cases, their primary goal is similar: optimizing resource usage by changing how data is stored and \python code is executed.

% -------------------------------------------------- %

\smallskip
\noindent
\textbf{\polars}~\cite{polars_github} is a \python library written in \rust and built on top of \textsf{arrow2}, the \rust implementation of the \arrow format.
% \footnote{\url{https://github.com/pola-rs/polars}}
The adoption of \arrow as the underlying data structure provides \polars with efficient data processing capabilities, optimized through parallel execution, cache-efficient algorithms, and efficient usage of resources.

\polars does not use an index, but each row is indexed by its integer position in the table.
Further, it has developed its own Domain Specific Language (DSL) for transforming data, whose core components are \textsf{Expressions} and \textsf{Contexts}.
% , which aims to be very usable and ensure human readability even for complex queries.
% Its core components are \textrm{Expressions} and \textrm{Contexts}.
\textsf{Expressions} facilitate concise and efficient data transformations, while \textsf{Contexts} categorize evaluations into three main types: filtering, grouping/aggregation, and selection.
% \emph{(i)} filtering,
% \emph{(ii)} grouping/aggregation,
% \emph{(iii)} selection.
\polars optimizes queries through early filters and projection pushdown.
Moreover, it supports both eager and lazy evaluation.
The lazy strategy allows to run queries in a streaming manner: instead of processing the entire data at once, they can run in batches, lightening the load on memory and CPU, hence allowing to process bigger datasets (even larger than memory) in a faster manner.

% -------------------------------------------------- %

\smallskip
\noindent
\textbf{\cudf}~\cite{cudf_github} is a component of the NVIDIA \textsf{RAPIDS} framework.
% \footnote{\url{https://github.com/rapidsai/cudf}}
Written in \textsf{C}/\textsf{C++} and \textsf{CUDA} (hence only compatible with NVIDIA GPUs), it offers a \pandas-like API to run general-purpose data science pipelines on GPUs, leveraging their computational power for accelerating data processing.
% which offers a \pandas-like API to run general-purpose data science pipelines on GPUs.
% Written in \textsf{C}/\textsf{C++} and \textsf{CUDA}, it harnesses the computational power of GPUs for accelerating data processing.

\cudf is built on top of \arrow and leverages parallelization to execute operations on different parts of columns simultaneously across all available GPU cores.
Note that \cudf uses a single GPU, while \textsf{Dask-CuDF} can be used for multi-GPU parallel computing.
Thus, it can perform efficient and high-performance computations, although it does not provide any optimization strategy for the execution of the pipelines.
% facilitates the interchange of tabular data across GPU processes.
% Further, it leverages parallelization to execute operations on different parts of columns simultaneously across GPU cores.
% This approach takes advantage of all available GPU cores, resulting in efficient and high-performance computations.
% Note that \cudf does not provide any optimization strategy for the execution of the pipelines.
% executes all operations on the GPU, but it

% -------------------------------------------------- %

\smallskip
\noindent
\textbf{\vaex}~\cite{vaex_github} is a \python library\footnote{Since the last commit on GitHub is dated about one year ago, the project might not be actively maintained at the moment.} designed to handle extremely large tabular datasets (such as astronomical catalogs), producing fast statistical analysis and visualization.
% \footnote{\url{https://github.com/vaexio/vaex}}
Supported by its \textsf{vaex-core} extension, written in \textsf{C}, \vaex achieves memory optimization through streaming algorithms, memory-mapped files, and a zero-copy policy, supporting the exploration of datasets that exceed the available memory.

\vaex enables efficient column-wise operations by wrapping a series of \numpy arrays as columns.
Only a small subset of \numpy functions benefits of lazy evaluation, storing their results as computation instructions and computing them only when needed.
% (e.g., \textrm{exp}, \textrm{cos}, \textrm{arcsin})
This is achieved through virtual columns, expressions that can be added to datasets without incurring additional memory usage.
\vaex can process random data subsets and export data in random order for efficient resource utilization.
Moreover, it also supports multithreading for faster computation on multi-core CPUs.
% By default, \vaex avoids unnecessary data copying and loading into memory.

% -------------------------------------------------- %

\smallskip
\noindent
\textbf{\datatable}~\cite{datatable_github} is a \python package that relies on the \textsf{Frame} object, similar to \pandas dataframes or \sql tables (i.e., data arranged in a two-dimensional array with rows and columns).
% \footnote{\url{https://github.com/h2oai/datatable}}
% While it shares similarities with \pandas, \datatable emphasizes speed and support for big data.

\datatable, which uses a native-\textsf{C} implementation for all data types, is developed to support column-oriented data storage, optimizing data access and processing.
\datatable enables memory-mapping of data on disk, allowing the seamless processing of out-of-memory datasets.
% , hence it is suitable for large-scale data analysis tasks.
\datatable incorporates multithreaded data processing, leveraging all available cores for time-consuming operations, and offers efficient algorithms for sorting, grouping, and joining operations.
To minimize unnecessary data copying, it adopts the copy-on-write technique for shared data, reducing memory overhead and improving the overall performance.

% -------------------------------------------------- %

\section{Bento}
\label{sec:dsandpipe}

In this section, we provide all details about the configuration of \bento, our framework for evaluating different dataframe libraries on data preparation pipelines.

% -------------------------------------------------- %

\subsubsection*{Datasets}

\begin{table}[!t]
\small
\centering
\caption{Features of the selected datasets.}
%\vspace{-0.1cm}
\label{tab:dataset}
\begin{adjustbox}{max width=\columnwidth}
\begin{tabular}{l|cccc|}
\cline{2-5}
                                            & \textbf{Athlete} & \textbf{Loan} & \textbf{Patrol} & \textbf{Taxi} \\ \hhline{-====}
\multicolumn{1}{|l|}{CSV Size (GB)}         & 0.03             & 1.6           & 6.7             & 10.9          \\ \hline
\multicolumn{1}{|l|}{\# Rows ($\times10^{6}$)} & 0.2              & 2             & 27              & 77            \\ \hline
\multicolumn{1}{|l|}{\# Columns}               & 15               & 151           & 34              & 18            \\ \hline
\multicolumn{1}{|l|}{\# Num - Str - Bool}     & 5-10-0                & 113-38-0           & 5-27-2   & 15-3-0            \\ \hline
\multicolumn{1}{|l|}{\% Null}                  &         9\%         & 31\%           &    22\%             &  0\%             \\ \hline
\multicolumn{1}{|l|}{Str Len Range}           & (1, 108)         & (1, 3988)     & (1, 2293)       & (1, 19)       \\ \hline
\end{tabular}

\end{adjustbox}  

\end{table}

For a comprehensive evaluation of the dataframe libraries on general, real-world data preparation use cases, we selected four datasets covering a variety of sizes, complexities, and features:
\textit{(i)} \athlete~\cite{kaggle_athlete_dataset}, collecting 0.2M records about the results achieved by the athletes through 120 years of Olympics;
% \footnote{\url{https://www.kaggle.com/datasets/heesoo37/120-years-of-olympic-history-athletes-and-results}}
\textit{(ii)} \loan~\cite{kaggle_loan_dataset}, containing 2M rows about loan applicants and their financial profiles from the LendingClub company;
% \footnote{\url{https://www.kaggle.com/datasets/wordsforthewise/lending-club}}
\textit{(iii)} \patrol~\cite{kaggle_patrol_dataset}, composed of 27M records about 11 years of traffic stops by California law enforcement agencies~\cite{pierson_2020_nhb_police_stops};
% \footnote{\url{https://www.kaggle.com/datasets/faressayah/stanford-open-policing-project}}
\textit{(iv)} \taxi~\cite{kaggle_taxi_dataset}, containing 77M records about taxi trips in New York City in 2015.
% \footnote{\url{https://www.kaggle.com/competitions/nyc-taxi-trip-duration}}
% , hence also suitable for evaluating scalability.
A summary of their main features is displayed in Table~\ref{tab:dataset}.
Note that we compute the percentage of missing values as the number of null cells over all cells in the dataframe.
%{\color{green}In this context, we define dataset sparsity considering both the percentage of missing values and the number of columns. A dataset is considered \emph{sparse} if it has a high percentage of missing values and a large number of columns (e.g., \loan). Conversely, a \emph{dense} dataset has few missing values and a relatively small number of columns (e.g., \taxi).}
Collected from Kaggle, all of these datasets had previously been adopted in related work about data preparation and dataframe libraries.
The three largest datasets were indeed used to evaluate the performance of \modin~\cite{petersohn2021modin@pvldb}, due to the variety of their features.
Instead, \athlete was exploited to analyze how different preparators are supported by commercial data preparation tools~\cite{hameed2020data_preparation_survey@sigmod_record}.

% -------------------------------------------------- %

\subsubsection*{Data Preparation Pipelines}
\label{subsec:dp_pipe}

% \usepackage{multirow}
% \usepackage{graphicx}
\begin{table*}[ht]
\small
\caption{Compatibility of dataframe libraries with Pandas API. 
(\dcmark) fully matches Pandas interface;
(\cmark) different interface;
($\circ $) missing from the API, but implemented by us to the best of our efforts.
}
\label{tab:api-compatibility}
%\resizebox{\textwidth}{!}{%
\begin{tabular}{r|r||c|c|c|c|c|c|c|}
\cline{2-9}
\textbf{} &
  \textbf{Preparator} &
  \textbf{SparkPD} &
  \textbf{SparkSQL} &
  \textbf{Modin} &
  \textbf{Polars} &
  \textbf{CuDF} &
  \textbf{Vaex} &
  \textbf{DataTable} \\ \hhline{-========}
\multicolumn{1}{|r|}{\multirow{2}{*}{\textbf{I/O}}} &
  load dataframe (\emph{{read}}) &
  \dcmark &
  \cmark &
  \dcmark &
  \dcmark &
  \dcmark &
  \cmark &
  {\color[HTML]{333333} \cmark} \\
\multicolumn{1}{|r|}{} &
  output dataframe (\emph{{write}}) &
  \dcmark &
  \cmark &
  \dcmark &
  \cmark &
  \dcmark &
  \cmark &
  {\color[HTML]{333333} \dcmark} \\ \hline
\multicolumn{1}{|r|}{\multirow{8}{*}{\textbf{EDA}}} &
  locate missing values (\emph{{isna}}) &
  \dcmark &
  $\circ $ &
  \dcmark &
  \cmark &
  \dcmark &
  $\circ $ &
  {\color[HTML]{333333} \cmark} \\
\multicolumn{1}{|r|}{} &
  locate outliers (\emph{{outlier}}) &
  \dcmark &
  \cmark &
  \dcmark &
  \dcmark &
  \dcmark &
  {\color[HTML]{333333} \cmark} &
  {\color[HTML]{333333} $\circ $} \\
\multicolumn{1}{|r|}{} &
  search by pattern (\emph{{srchptn}}) &
  \dcmark &
  \cmark &
  \dcmark &
  \dcmark &
  \dcmark &
  \dcmark &
  {\color[HTML]{333333} \dcmark} \\
\multicolumn{1}{|r|}{} &
  sort values (\emph{{sort}}) &
  \dcmark &
  \dcmark &
  \dcmark &
  \dcmark &
  \dcmark &
  \dcmark &
  {\color[HTML]{333333} \dcmark} \\
\multicolumn{1}{|r|}{} &
  get columns list (\emph{{getcols}}) &
  \dcmark &
  \dcmark &
  \dcmark &
  \dcmark &
  \dcmark &
  {\color[HTML]{333333} \cmark} &
  {\color[HTML]{333333} \cmark} \\
\multicolumn{1}{|r|}{} &
  get columns types (\emph{{dtypes}}) &
  \dcmark &
  \dcmark &
  \dcmark &
  \dcmark &
  \dcmark &
  \dcmark &
  {\color[HTML]{333333} \cmark} \\
\multicolumn{1}{|r|}{} &
  get dataframe statistics (\emph{{stats}}) &
  \dcmark &
  \dcmark &
  \dcmark &
  \dcmark &
  \dcmark &
  \dcmark &
  {\color[HTML]{333333} $\circ $} \\
\multicolumn{1}{|r|}{} &
  query columns (\emph{{query}}) &
  \dcmark &
  \cmark &
  \dcmark &
  \dcmark &
  \dcmark &
  {\color[HTML]{333333} \cmark} &
  {\color[HTML]{333333} $\circ $} \\ \hline
\multicolumn{1}{|r|}{\multirow{9}{*}{\textbf{DT}}} &
  cast columns types (\emph{{cast}}) &
  \dcmark &
  \cmark &
  \dcmark &
  {\color[HTML]{333333} \cmark} &
  \dcmark &
  \dcmark &
  {\color[HTML]{333333} $\circ $} \\
\multicolumn{1}{|r|}{} &
  delete columns (\emph{{drop}}) &
  \dcmark &
  \dcmark &
  \dcmark &
  \dcmark &
  \dcmark &
  $\circ $ &
  {\color[HTML]{333333} $\circ $} \\
\multicolumn{1}{|r|}{} &
  rename columns (\emph{{rename}}) &
  \dcmark &
  $\circ $ &
  \dcmark &
  \dcmark &
  \dcmark &
  \dcmark &
  {\color[HTML]{333333} $\circ $} \\
\multicolumn{1}{|r|}{} &
  pivot table (\emph{{pivot}}) &
  \dcmark &
  {\color[HTML]{333333} \cmark} &
  \dcmark &
  {\color[HTML]{333333} \cmark} &
  \dcmark &
  $\circ $ &
  {\color[HTML]{333333} $\circ $} \\
\multicolumn{1}{|r|}{} &
  calculate column using expressions (\emph{{calccol}}) &
  \dcmark &
  $\circ $ &
  \dcmark &
  \dcmark &
  {\color[HTML]{333333} $\circ $} &
  \dcmark &
  {\color[HTML]{333333} $\circ $} \\
\multicolumn{1}{|r|}{} &
  join dataframes (\emph{{join}}) &
  \dcmark &
  $\circ $ &
  \dcmark &
  {\color[HTML]{333333} \cmark} &
  \dcmark &
  $\circ $ &
  {\color[HTML]{333333} $\circ $} \\
\multicolumn{1}{|r|}{} &
  one hot encoding (\emph{{onehot}}) &
  \dcmark &
  $\circ $ &
  \dcmark &
  \dcmark &
  \dcmark &
  \cmark &
  {\color[HTML]{333333} $\circ $} \\
\multicolumn{1}{|r|}{} &
  categorical encoding (\emph{{catenc}}) &
  \dcmark &
  \cmark &
  \dcmark &
  {\color[HTML]{333333} \cmark} &
  \dcmark &
  \cmark &
  {\color[HTML]{333333} $\circ $} \\
\multicolumn{1}{|r|}{} &
  group dataframe (\emph{{group}}) &
  \dcmark &
  \cmark &
  \dcmark &
  \dcmark &
  \dcmark &
  \dcmark &
  {\color[HTML]{333333} \dcmark} \\ \hline
\multicolumn{1}{|r|}{\multirow{8}{*}{\textbf{DC}}} &
  change date \& time format (\emph{{chdate}}) &
  \dcmark &
  \cmark &
  \dcmark &
  $\circ $ &
  \dcmark &
  {\color[HTML]{333333} $\circ $} &
  {\color[HTML]{333333} $\circ $} \\
\multicolumn{1}{|r|}{} &
  delete empty and invalid rows (\emph{{dropna}}) &
  \dcmark &
  \cmark &
  \dcmark &
  \cmark &
  \dcmark &
  \dcmark &
  {\color[HTML]{333333} $\circ $} \\
\multicolumn{1}{|r|}{} &
  set content case (\emph{{setcase}}) &
  \dcmark &
  \cmark &
  \dcmark &
  \cmark &
  \dcmark &
  \dcmark &
  {\color[HTML]{333333} \dcmark} \\
\multicolumn{1}{|r|}{} &
  normalize numeric values (\emph{{norm}}) &
  \dcmark &
  \cmark &
  \dcmark &
  \dcmark &
  \dcmark &
  \dcmark &
  {\color[HTML]{333333} $\circ $} \\
\multicolumn{1}{|r|}{} &
  deduplicate rows (\emph{{dedup}}) &
  \dcmark &
  \cmark &
  \dcmark &
  {\color[HTML]{333333} \cmark} &
  \dcmark &
  $\circ $ &
  {\color[HTML]{333333} $\circ $} \\
\multicolumn{1}{|r|}{} &
  fill empty cells (\emph{{fillna}}) &
  \dcmark &
  \cmark &
  \dcmark &
  $\circ $ &
  \dcmark &
  \dcmark &
  {\color[HTML]{333333} $\circ $} \\
\multicolumn{1}{|r|}{} &
  replace values occurrences (\emph{{replace}}) &
  \dcmark &
  {\color[HTML]{333333} \cmark} &
  \dcmark &
  $\circ $ &
  \dcmark &
  {\color[HTML]{333333} \cmark} &
  {\color[HTML]{333333} $\circ $} \\
\multicolumn{1}{|r|}{} &
  edit \& replace cell data (\emph{{edit}}) &
  \dcmark &
  $\circ $ &
  \dcmark &
  {\color[HTML]{333333} \cmark} &
  {\color[HTML]{333333} \dcmark} &
  \dcmark &
  {\color[HTML]{333333} \dcmark} \\ \hline
\end{tabular}%
%}
\end{table*}

For each dataset, we selected three notebooks among the top-voted Kaggle entries that proposed a non-trivial solution for the task at hand (i.e., Kaggle competitions) and extracted its data preparation pipeline---typically the first part of the notebook, which precedes the definition of a machine learning model.

Overall, the pipelines employ a set of 27 preparators (reported in Table~\ref{tab:api-compatibility}) that can be clustered into four main stages:
\begin{itemize}[]
    \item \emph{Input/output} (I/O): preparators to handle the input/output of data in various formats, such as databases, \csv or \parquet files, Web APIs, etc.
    \item \emph{Exploratory data analysis} (EDA): preparators to support the exploratory analysis of data features, to better understand the data at hand and detect errors or anomalies.
    \item \emph{Data transformation} (DT): preparators for transforming the data to make it more suitable for analysis, such as categorical encoding, join, aggregation, etc.
    \item \emph{Data cleaning} (DC): preparators to improve the quality of the data, hence the reliability of the results of its analysis (e.g., handling missing or erroneous data and correcting or removing outliers).
\end{itemize}

% -------------------------------------------------- %

\subsubsection*{Pipeline Specification}

\bento is a \python framework that provides a general shared interface to design the data preparation pipeline, ignoring the differences in the APIs of the considered libraries.
For preparator names, we adopted the convention proposed by Hameed and Naumann~\cite{hameed2020data_preparation_survey@sigmod_record}.
Each pipeline can be defined through a \textsf{JSON} file by specifying the sequence of preparators to employ.
Our tool automatically deploys the pipeline for every specific library.

As shown in Table~\ref{tab:api-compatibility}, not all preparators are available in every dataframe library.
In such cases, we implemented them with our best effort to avoid the transition to \pandas and back.
When a preparator is directly available in the API of the library, we denote whether its interface fully aligns with \pandas ($\dcmark$) or differs in the adopted name ($\cmark$).

Unless differently stated, we load data from \csv files and measure the execution time under three key settings:
\emph{(i)} \emph{function-core}, which evaluates each preparator alone (if the library adopts lazy evaluation, it requires to force the execution after each API call);
\emph{(ii)} \emph{pipeline-stage}, which measures the runtime for the execution of each of the four stages (i.e., I/O, EDA, DT, and DC);
\emph{(iii)} \emph{pipeline-full}, which measures the runtime for the execution of the entire pipeline.
Finally, our framework also supports the execution of the pipeline in \docker containers, which allow to specify core and memory usage and isolate \python environments to prevent package conflicts.

% -------------------------------------------------- %

\subsubsection*{Hardware and Software}

All measurements are obtained as the average value over ten runs on a machine equipped with two AMD EPYC Rome 7402 CPUs (48 threads running at 2.8 GHz) and 512 GB of RAM, using \python 3.9, when not specified otherwise.
The machine also features a NVIDIA A100 GPU with 40 GB of RAM and CUDA 12.1.
The \ray engine was configured with default settings, resulting in 48 worker threads running.
The \dask engine was configured comparably, leading to the creation of 8 workers and 48 threads for each execution.

To ensure accuracy and avoid warm-up overhead, the assessment of execution time occurs once the system (e.g., \textsf{JVM}) has completed its warm-up process.
To assess the performance of the libraries across machines with different hardware specifications, we simulated three distinct machine configurations, as detailed in Table~\ref{tab:machineconfiguration}.
Finally, we evaluated scalability on incremental samples of \taxi and \patrol.

% -------------------------------------------------- %

\section{Evaluation Results}
\label{sec:results}

In this section, we report and analyze the results of the extensive experimental evaluation of the presented dataframe libraries.
Our main goal is to provide data scientists and practitiones with useful insights for supporting them in the selection of the best solution for their data preparation tasks.
Therefore, our comparison is designed to assess the performance of dataframe libraries based on the operations to carry out (considering both the distinct preparators and the benefits introduced by lazy evaluation, when supported), the size and the features of the dataset at hand, and the configuration of the machine on which the pipeline is executed.

\begin{figure*}[ht]

\begin{subfigure}{\textwidth}
         \centering
         \hspace*{0.8cm}\includegraphics[width=0.945\linewidth]{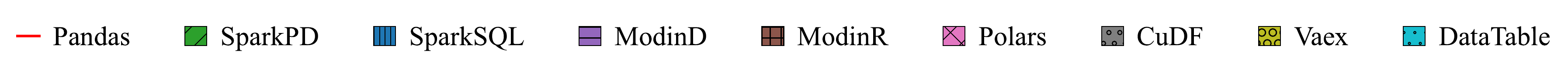}
\end{subfigure}

\begin{subfigure}{\textwidth}
    \begin{subfigure}[b]{0.245\textwidth}
         \centering
         \includegraphics[width=\textwidth]{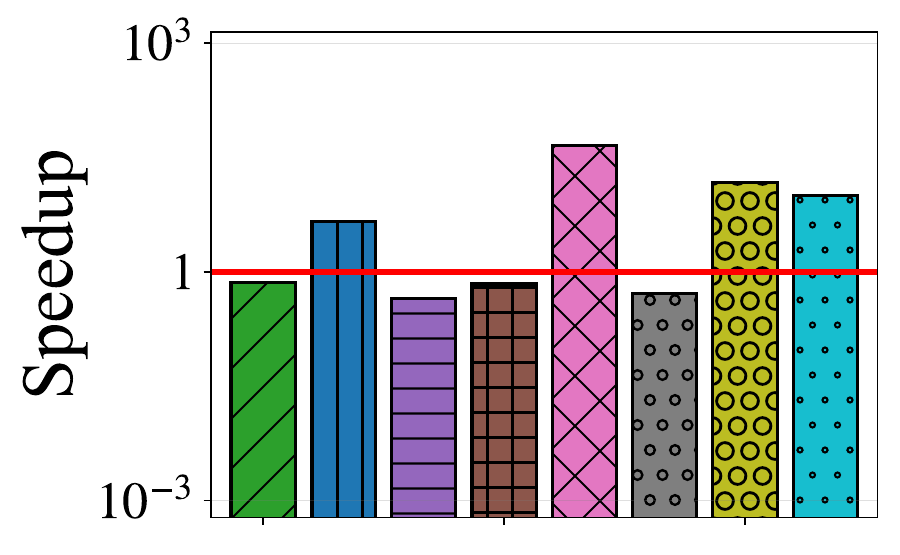}
         \vspace*{-0.5cm}
         \caption{Athlete, EDA}
         \label{fig:athlete_eda}
         \vspace{0.25cm}
    \end{subfigure}
    \hfill
        \begin{subfigure}[b]{0.245\textwidth}
         \centering
         \includegraphics[width=\textwidth]{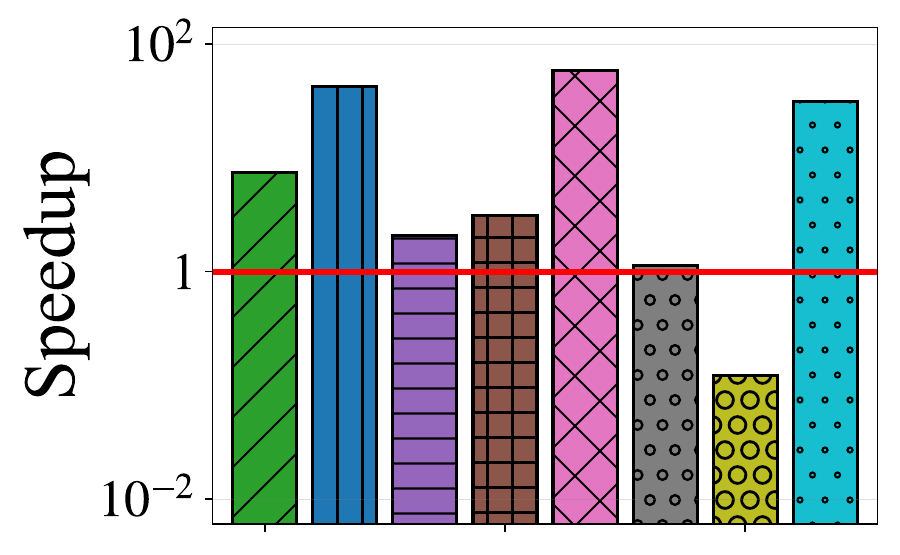}
         \vspace*{-0.5cm}
         \caption{Loan, EDA}
         \label{fig:loan_eda}
         \vspace{0.25cm}
    \end{subfigure}
    \hfill
        \begin{subfigure}[b]{0.25\textwidth}
         \centering
         \includegraphics[width=\textwidth]{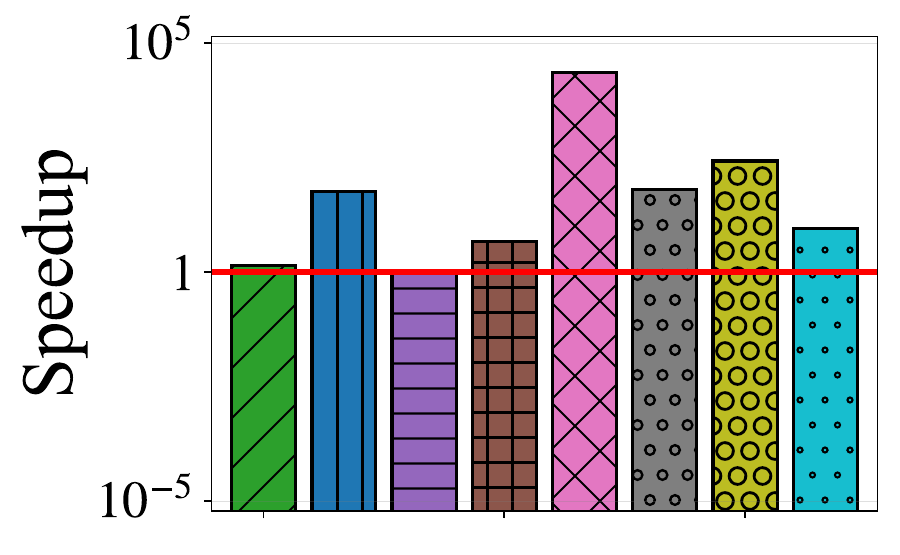}
         \vspace*{-0.55cm}
         \caption{Patrol, EDA}
         \label{fig:patrol_eda}
         \vspace{0.25cm}
    \end{subfigure}
    \hfill
        \begin{subfigure}[b]{0.245\textwidth}
         \centering
         \includegraphics[width=\textwidth]{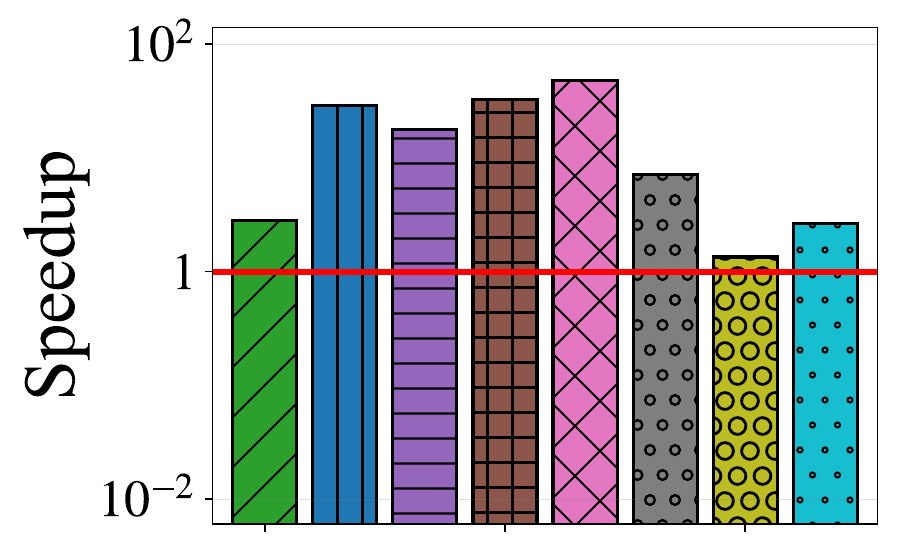}
         \vspace*{-0.5cm}
         \caption{Taxi, EDA}
         \label{fig:taxi_eda}
         \vspace{0.25cm}
    \end{subfigure}
\end{subfigure}
\hfill
\begin{subfigure}{\textwidth}
    \begin{subfigure}[b]{0.245\textwidth}
         \centering
         \includegraphics[width=\textwidth]{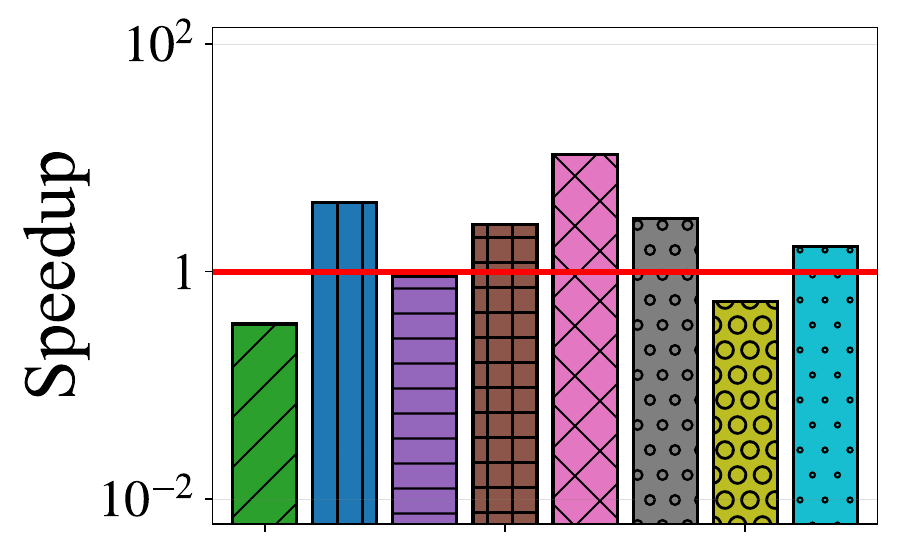}
         \vspace*{-0.5cm}
         \caption{Athlete, DT}
         \label{fig:athlete_dt}
         \vspace{0.25cm}
    \end{subfigure}
    \hfill
        \begin{subfigure}[b]{0.245\textwidth}
         \centering
         \includegraphics[width=\textwidth]{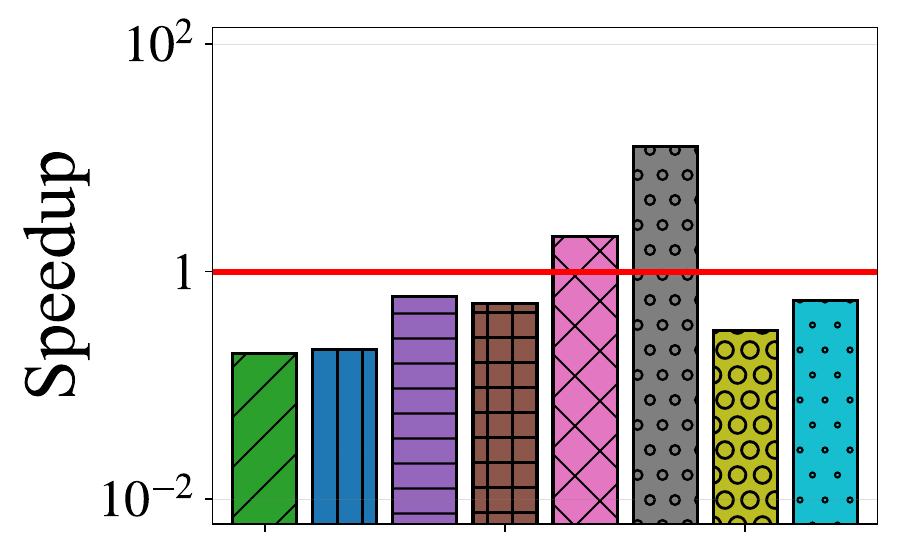}
         \vspace*{-0.5cm}
         \caption{Loan, DT}
         \label{fig:loan_dt}
         \vspace{0.25cm}
    \end{subfigure}
    \hfill
        \begin{subfigure}[b]{0.25\textwidth}
         \centering
         \includegraphics[width=\textwidth]{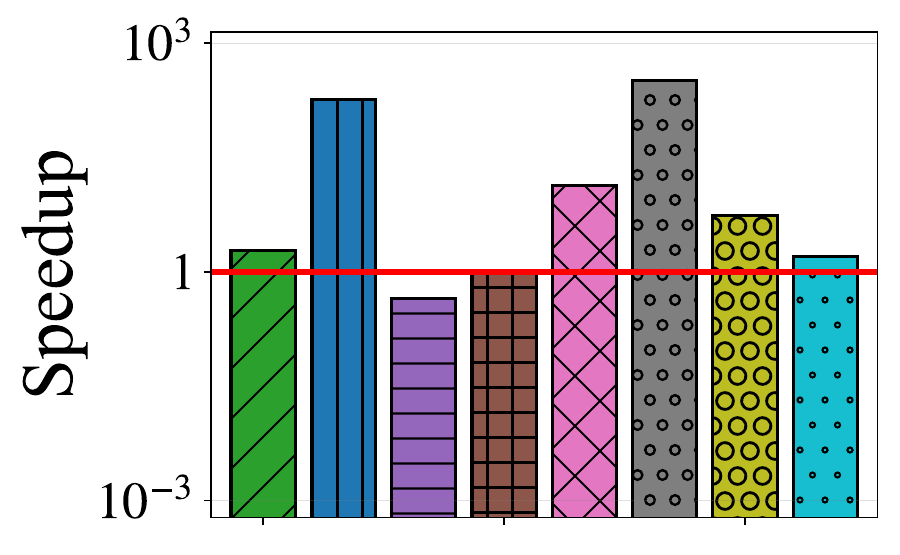}
         \vspace*{-0.525cm}
         \caption{Patrol, DT}
         \label{fig:patrol_dt}
         \vspace{0.25cm}
    \end{subfigure}
    \hfill
        \begin{subfigure}[b]{0.245\textwidth}
         \centering
         \includegraphics[width=\textwidth]{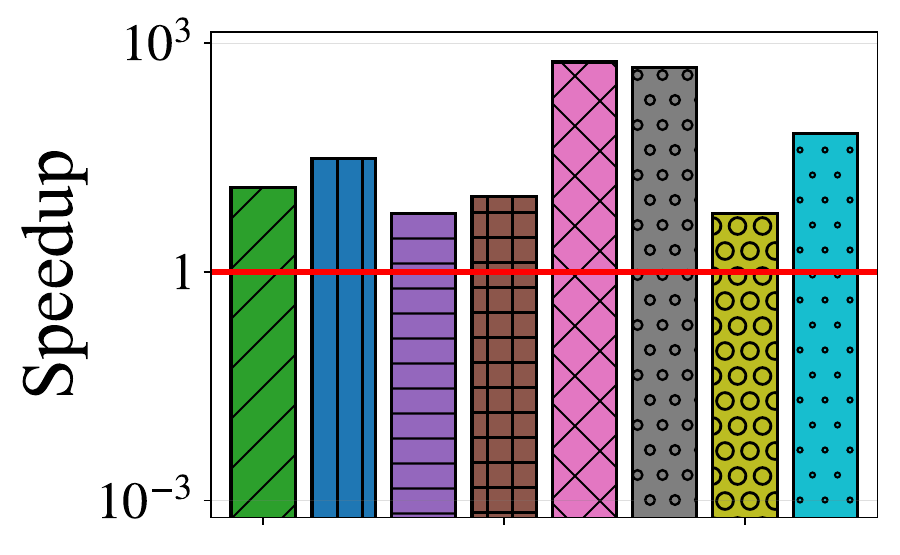}
         \vspace*{-0.525cm}
         \caption{Taxi, DT}
         \label{fig:taxi_dt}
         \vspace{0.25cm}
    \end{subfigure}
\end{subfigure}
\hfill
\begin{subfigure}{\textwidth}
    \begin{subfigure}[b]{0.245\textwidth}
         \centering
         \includegraphics[width=\textwidth]{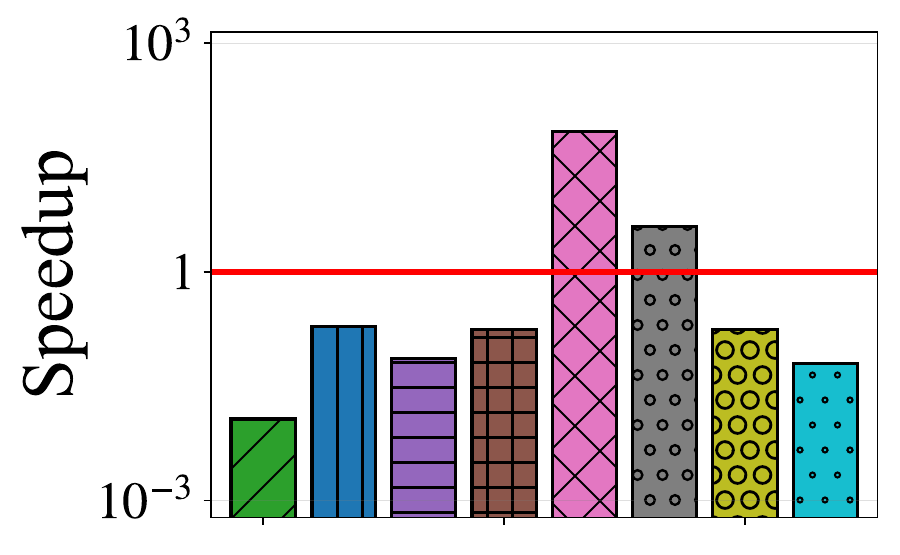}
         \vspace*{-0.5cm}
         \caption{Athlete, DC}
         \label{fig:athlete_dc}
         \vspace{0.25cm}
    \end{subfigure}
    \hfill
        \begin{subfigure}[b]{0.245\textwidth}
         \centering
         \includegraphics[width=\textwidth]{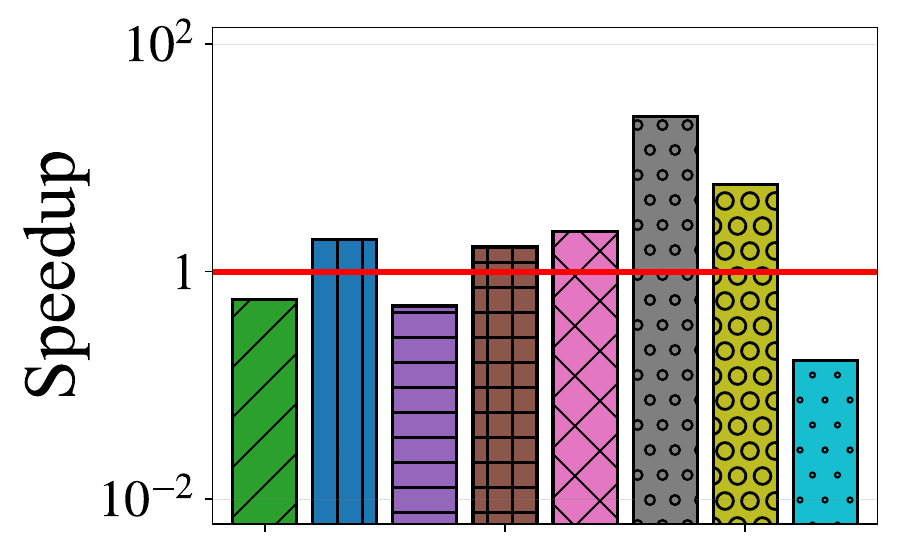}
         \vspace*{-0.5cm}
         \caption{Loan, DC}
         \label{fig:loan_dc}
         \vspace{0.25cm}
    \end{subfigure}
    \hfill
        \begin{subfigure}[b]{0.25\textwidth}
         \centering
         \includegraphics[width=\textwidth]{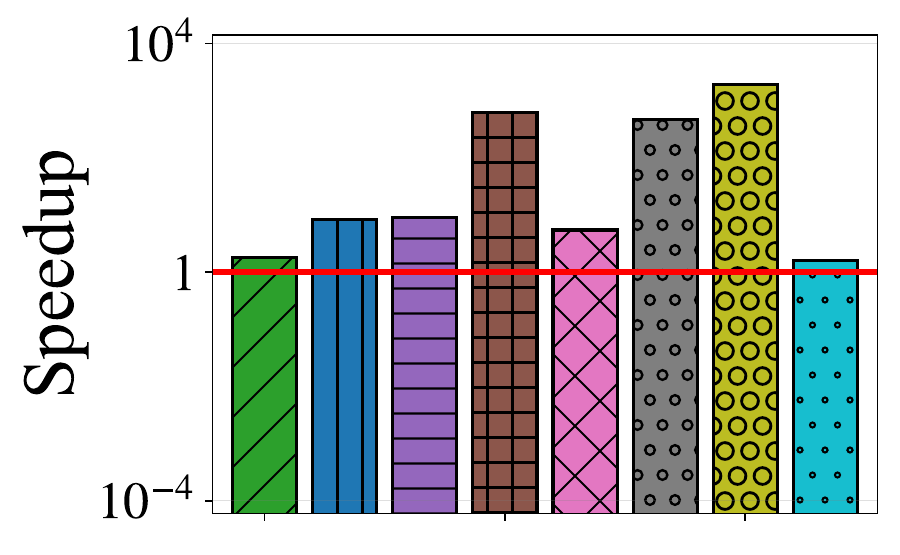}
         \vspace*{-0.55cm}
         \caption{Patrol, DC}
         \label{fig:patrol_dc}
         \vspace{0.25cm}
    \end{subfigure}
    \hfill
        \begin{subfigure}[b]{0.245\textwidth}
         \centering
         \includegraphics[width=\textwidth]{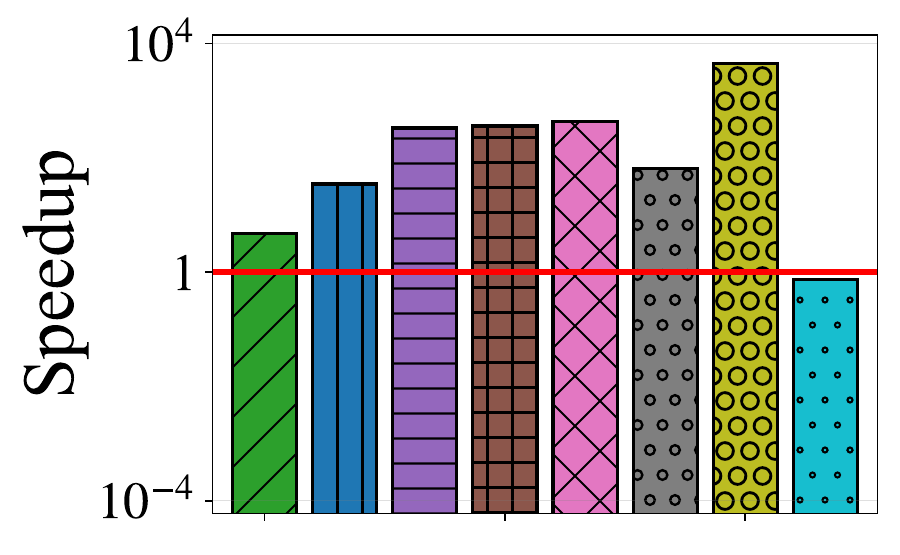}
         \vspace*{-0.55cm}
         \caption{Taxi, DC}
         \label{fig:taxi_dc}
         \vspace{0.25cm}
    \end{subfigure}
\end{subfigure}
\caption{Average speedup over Pandas computed for each stage (EDA, DT, DC) on the three data preparation pipelines.}
\label{fig:avg_stage}
\end{figure*}

In particular, we organize our evaluation in multiple subsections, each designed to answer one of the following research questions:

\begin{enumerate}[label=Q\arabic{enumi}.,ref=Step \arabic{enumi},leftmargin=3em]
    \item What is the performance of the dataframe libraries in running data preparation pipelines on datasets of different size and features? (Section~\ref{subsec:pipeline_performance})
    \smallskip
    \item How does lazy evaluation impact on the performance of the libraries that support it? 
    (Section~\ref{sec:evalstrategies}) 
    \smallskip
    \item How do libraries scale by varying the size of the dataset and the configuration of the underlying machine (i.e., from laptop to server)? (Section~\ref{sec:scalability})
    \smallskip
    \item How do libraries perform on the standard queries of the TPC-H benchmark? (Section~\ref{sec:tpch})
    \smallskip
\end{enumerate}

% -------------------------------------------------- %
%\newpage
\subsection{Evaluation on Data Preparation Pipelines} \label{subsec:pipeline_performance}
\begin{figure*}[ht]
\begin{subfigure}{\textwidth}
         \centering
         \hspace*{0.9cm}\includegraphics[width=0.94\linewidth]{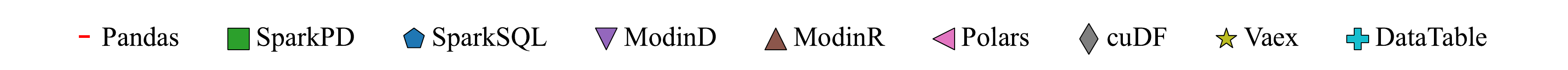}
\end{subfigure}
\hfill
\begin{subfigure}{\textwidth}
         \centering
         \includegraphics[width=\linewidth]{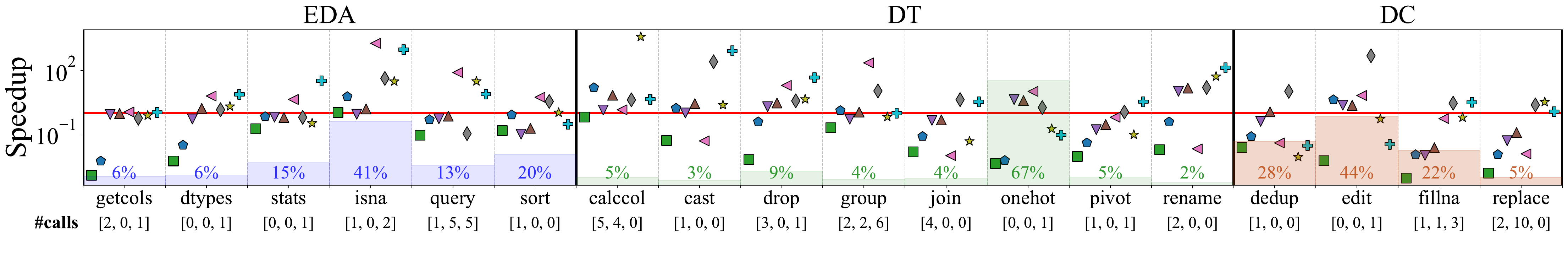}
         \vspace*{-0.6cm}
         \caption{Athlete}
         \label{fig:athlete_core}
         \vspace{10pt}
\end{subfigure}
\hfill
\begin{subfigure}{\textwidth}
         \centering
         \includegraphics[width=\linewidth]{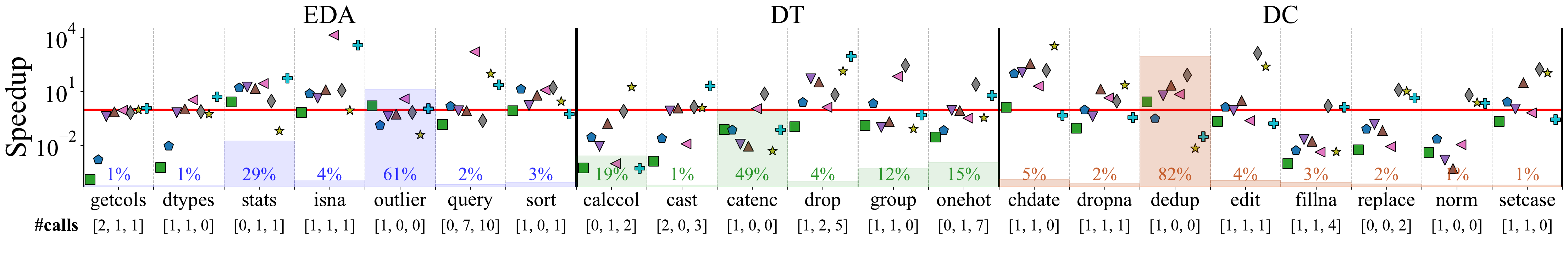}
         \vspace*{-0.6cm}
         \caption{Loan}
         \label{fig:loan_core}
         \vspace{10pt}
\end{subfigure}
\hfill
\begin{subfigure}{\textwidth}
         \centering
         \includegraphics[width=\linewidth]{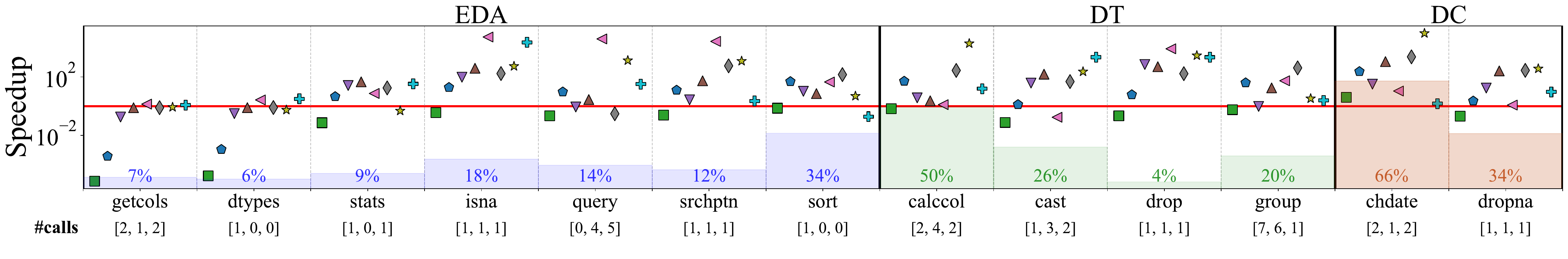}
         \vspace*{-0.6cm}
         \caption{Patrol}
         \label{fig:patrol_core}
         \vspace{10pt}
\end{subfigure}
\hfill
\begin{subfigure}{\textwidth}
         \centering
          \includegraphics[width=\linewidth]{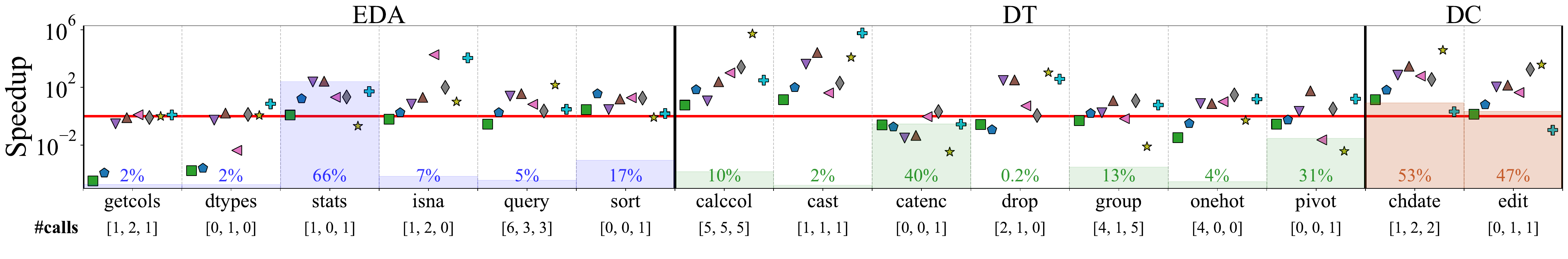}
         \vspace*{-0.6cm}
         \caption{Taxi}
         \label{fig:nyc_core}
\end{subfigure}
%\vspace*{-0.6cm}
\caption{Average speedup over \pandas computed for each preparator of the three data preparation pipelines per datasets.
For each preparator, we report the number of calls in each pipeline and its average impact on the stage execution time.}
\label{fig:speedup}
\end{figure*}

\begin{figure*}[ht]

\begin{subfigure}{\textwidth}
         \centering
         \hspace*{0.8cm}\includegraphics[width=0.945\linewidth]{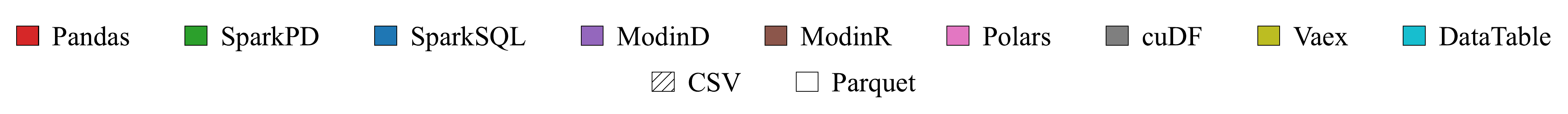}
\end{subfigure}

\begin{subfigure}{\textwidth}
    \begin{subfigure}[b]{0.245\textwidth}
         \centering
         \includegraphics[width=\textwidth]{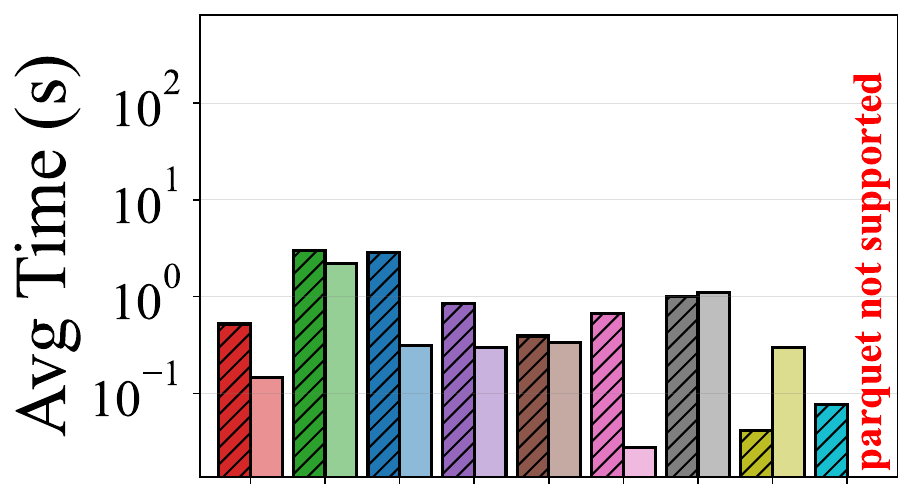}
         \vspace*{-0.4cm}
         \caption{Athlete, Read}
         \label{fig:athlete_read}
    \end{subfigure}
    \hfill
    \begin{subfigure}[b]{0.245\textwidth}
         \centering
         \includegraphics[width=\textwidth]{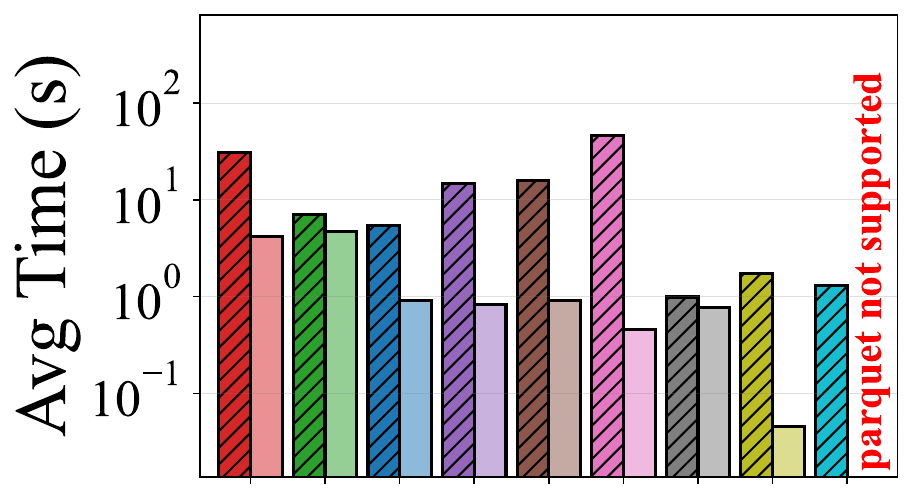}
         \vspace*{-0.4cm}
         \caption{Loan, Read}
         \label{fig:loan_read}
    \end{subfigure}
    \hfill
    \begin{subfigure}[b]{0.245\textwidth}
         \centering
         \includegraphics[width=\textwidth]{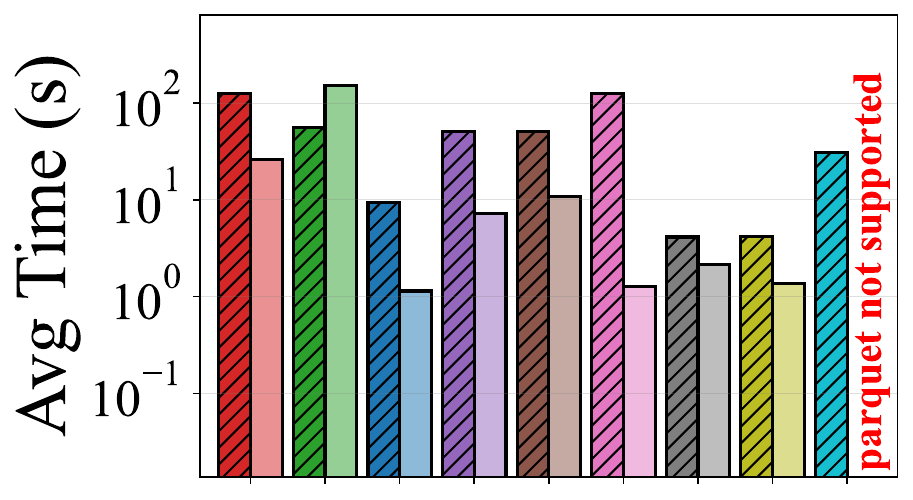}
         \vspace*{-0.41cm}
         \caption{Patrol, Read}
         \label{fig:patrol_read}
    \end{subfigure}
    \hfill
    \begin{subfigure}[b]{0.245\textwidth}
         \centering
         \includegraphics[width=\textwidth]{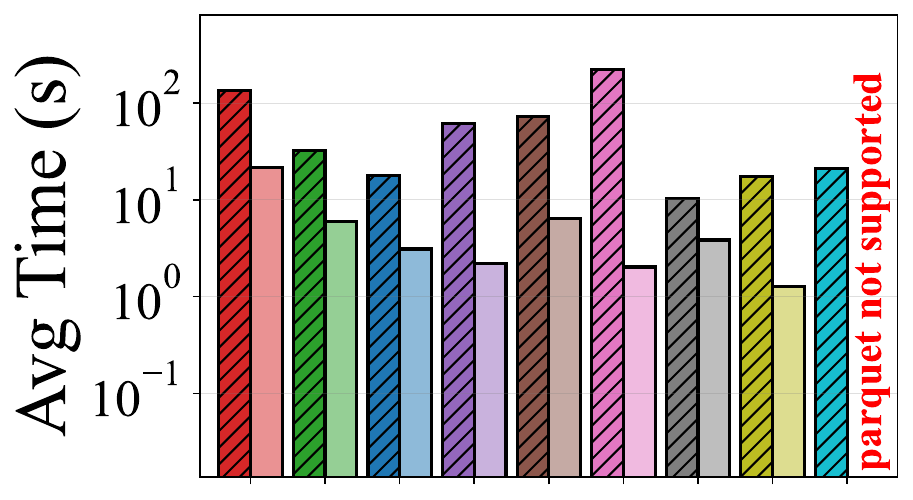}
         \vspace*{-0.4cm}
         \caption{Taxi, Read}
         \label{fig:taxi_read}
    \end{subfigure}
\end{subfigure}
%\vspace{-0.4cm}
\caption{Average runtime for reading \csv and \parquet files (lower is better).}
\label{fig:read_csv_parq}
\vspace{15pt}
\begin{subfigure}{\textwidth}
    \begin{subfigure}[b]{0.245\textwidth}
         \centering
         \includegraphics[width=\textwidth]{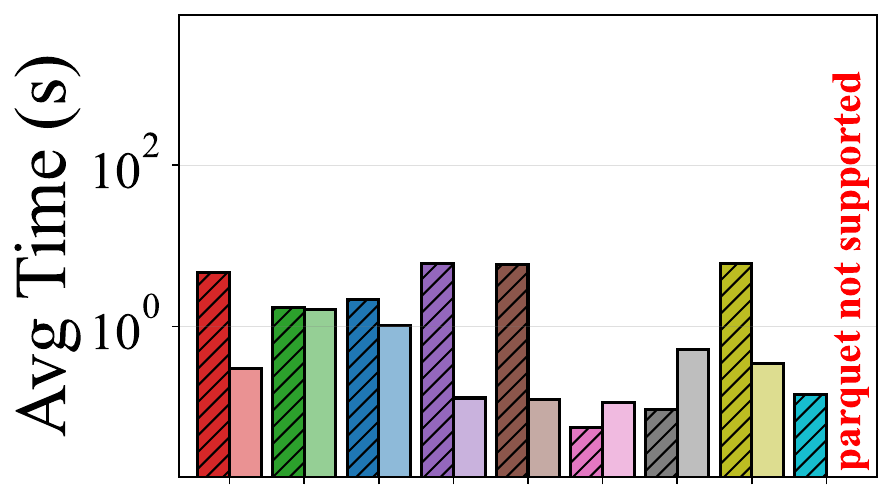}
         \vspace*{-0.4cm}
         \caption{Athlete, Write}
         \label{fig:athlete_write}
    \end{subfigure}
    \hfill
        \begin{subfigure}[b]{0.245\textwidth}
         \centering
         \includegraphics[width=\textwidth]{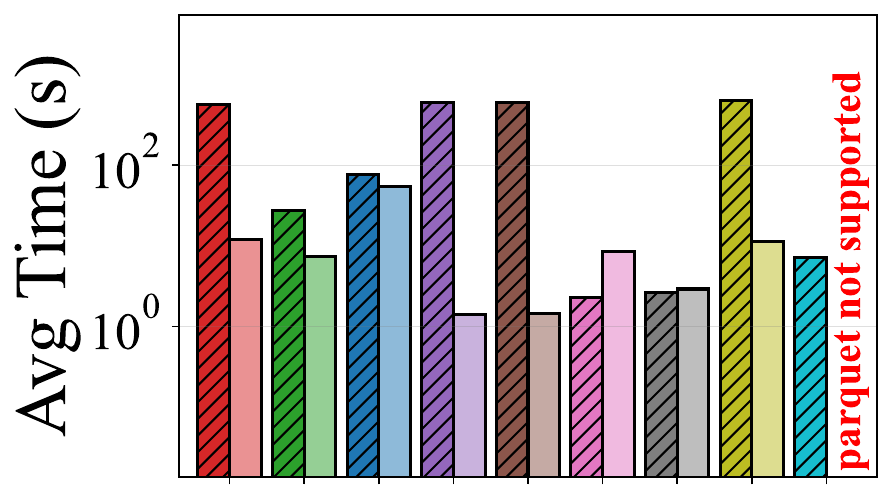}
         \vspace*{-0.4cm}
         \caption{Loan, Write}
         \label{fig:loan_write}
    \end{subfigure}
    \hfill
        \begin{subfigure}[b]{0.245\textwidth}
         \centering
         \includegraphics[width=\textwidth]{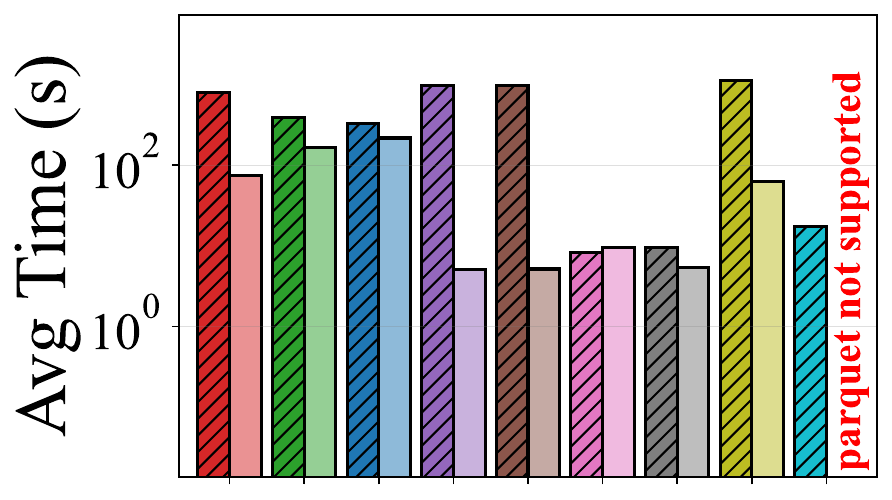}
         \vspace*{-0.41cm}
         \caption{Patrol, Write}
         \label{fig:patrol_write}
    \end{subfigure}
    \hfill
        \begin{subfigure}[b]{0.245\textwidth}
         \centering
         \includegraphics[width=\textwidth]{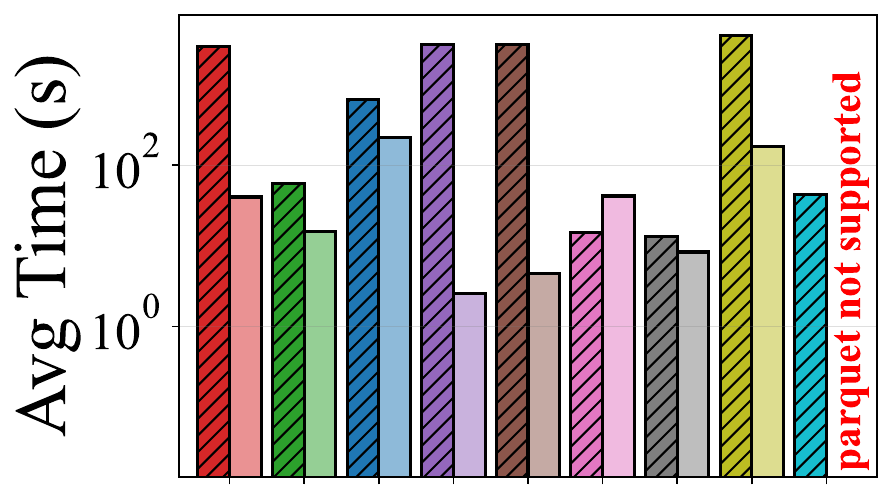}
         \vspace*{-0.4cm}
         \caption{Taxi, Write}
         \label{fig:taxi_write}
    \end{subfigure}
\end{subfigure}
%\vspace{-0.4cm}
\caption{Average runtime for writing \csv and \parquet files (lower is better).}
\label{fig:write_csv_parq}

\end{figure*}

\smallskip
\begin{mdframed}
    \textit{Summary---}For EDA, \polars is generally the best performer.
    For DT, if a GPU is available, \cudf generally outperforms other libraries.
    For DC, \vaex achieves notable results on the largest datasets.
    % otherwise, \sparksql is consistently fast.
    Finally, \cudf and \polars appear to be the best choices to read and write \csv files, respectively.
    %when working with \csv files  excels in reading, while \polars is the best choice for writing.}
\end{mdframed}
\medskip
For each of the selected datasets (\athlete, \loan, \patrol, \taxi), Figure~\ref{fig:avg_stage} and Figure~\ref{fig:speedup} show the average speedup over \pandas achieved by its alternatives in the execution of the three data preparation pipelines per dataset, focusing on the stages of \textrm{EDA}, \textrm{DT}, and \textrm{DC}.
In particular, Figure~\ref{fig:avg_stage} considers each stage in its entirety, allowing to perform lazy evaluation at the stage level when supported, Figure~\ref{fig:speedup} shows the performance separately for each distinct preparator (i.e., we force the execution for each of them).
The \textrm{I/O} stage is considered separately in Figures~\ref{fig:read_csv_parq} and~\ref{fig:write_csv_parq}.
The bars in Figure~\ref{fig:avg_stage} and the markers in Figure~\ref{fig:speedup} denote the speedup achieved over \pandas (the red line), defined as follows:

\medskip
\[\textit{\textrm{speedup}} = \frac{Time{\langle \textit{\textrm{Pandas}}, \textit{\textrm{prep/stage}} \rangle}}{Time{\langle \textit{\textrm{lib}}, \textit{\textrm{prep/stage}}} \rangle}\]
\medskip

\noindent where $Time{\langle \textit{\textrm{lib}}, \textit{\textrm{prep/stage}} \rangle}$ is the time required by the library \textit{\textrm{lib}} to run the considered preparator/stage.
Thus, a value above (below) the red line denotes that the library outperforms (fall behind) \pandas.
%The logarithmic scale enables to easily assess the order of magnitude of the improvement (if any) of each library over \pandas.
Further, Figure~\ref{fig:speedup} shows for each preparator the number of calls in each of the three pipelines and the impact on its stage, depicted by the bars in the background and defined as a percentage as follows:

\medskip
\[\textit{\textrm{Impact}} = \frac{Time{\langle \textit{\textrm{dataset}}, \textit{\textrm{prep}} \rangle}}{Time{\langle \textit{\textrm{dataset}}, \textit{\textrm{stage}} \rangle}}\times 100\]
\medskip

\noindent where $Time{\langle \textit{\textrm{dataset}}, \textit{\textrm{prep}} \rangle}$ is the average time for each stage preparator and $Time{\langle \textit{\textrm{dataset}}, \textit{\textrm{stage}} \rangle}$ is the sum of the average times for all preparators of the stage on the three dataset pipelines\footnote{To minimize bias from particularly fast or slow libraries, both in the numerator and in the denominator time values below the 20th percentile and above the 80th percentile were excluded.}.

In the next subsections, we analyze in detail the results obtained for each of the main data preparation stages, considering both the single preparators and the entire stages.
% Note that not all preparators have the same impact on the pipeline, but their weight depends on the time required for their execution (as we point out when analyzing the related results).
% At the end we analyze, if there is difference between real-world findings and the result of libraries on the TPC-H benchmark.

% -------------------------------------------------- %
% \input{input/images/tab5/tab5}

% \begin{figure*}[ht]
% \begin{subfigure}{\textwidth}
%          \centering
%          \hspace*{0.9cm}\includegraphics[width=0.94\linewidth]{input/images/tab4/legend.pdf}
% \end{subfigure}
% \hfill
% \begin{subfigure}{\textwidth}
%          \centering
%          \includegraphics[width=\linewidth]{input/images/tab4/state_patrol.pdf}
%          \caption{Patrol}
%          \label{fig:patrol_core}
% \end{subfigure}
% \hfill
% \begin{subfigure}{\textwidth}
%          \centering
%           \includegraphics[width=\linewidth]{input/images/tab4/nyc_taxi.pdf}
%          \caption{Taxi}
%          \label{fig:nyc_core}
% \end{subfigure}
% \caption{Speedup over \pandas for each preparator on \patrol and \taxi.}
% \label{fig:core_sp}
% \end{figure*}

\subsubsection*{Explorative Data Analysis (EDA)}

For EDA, \polars clearly stands out as the best performer on all datasets (Figures~\ref{fig:athlete_eda}-\ref{fig:taxi_eda}).
As depicted in Figure~\ref{fig:speedup}, \polars always registers the best performance for \emph{isna} and \emph{outlier}, the preparators with the greatest impact on \athlete and \loan, respectively.
For the former preparator, \polars is up to 10,000 times faster than \pandas and 3 times faster than the second-best performer, i.e., \datatable.
Both libraries avoid element-wise comparisons by using a special encoding to efficiently track null values: while \polars relies on the validity bitmap used by most types of \arrow arrays in their metadata~\cite{polars_null_values}, \datatable encodes them with sentinel values~\cite{datatable_github}.
For the latter preparator, outliers are located by filtering the dataframe using the values returned by the \emph{quantile} function.
\pandas relies on \numpy exact quantile computation (involving sorting and interpolation)~\cite{pandas_quantile}, while \polars and \spark employ approximate methods that avoid sorting~\cite{pyspark_quantile}.
The performance gap between \sparksql and \sparkpd arises from differences in their underlying implementations.
%(with \sparksql suffering additional overhead from its \sql engine compared to \sparkpd).

Although \vaex tends to be very efficient in column-wise operations (e.g, \emph{srchptn}) and can handle filtering well by internally tracking selected rows without copying data, operations based on percentiles are penalized by the complexity of the required calculations (i.e., determining min/max column values, cumulative sums, and grid interpolation)~\cite{vaex_api}.

For \emph{sort}, \cudf and \polars are the fastest libraries overall.
While \cudf leverages on high-performance \textsf{C++} parallel algorithms from the \textsf{Thrust} library~\cite{thrust}, \polars achieves high performance through its efficient multi-threaded \rust implementation~\cite{polars_sort}.
\sparksql shows remarkable advantages over \pandas as the datasets grow larger, while \sparkpd performs similarly to the baseline.
\modin performs significantly worse than \pandas on \athlete as it requires to apply \pandas implementation of the sorting preparator within each data partition~\cite{modin_architecture}, introducing significant latency to partition and merge results.
For \emph{stats}, \datatable outperforms \pandas by up to 50 times by computing statistics either during the creation of the \textsf{Frame} object or efficiently on-demand when needed, while \modin (with both \ray and \dask) is up to 200 times faster than the baseline on \taxi thanks to its multi-threading capabilities.

\vaex achieves notable performance for \emph{query}, for which it excels on the \taxi dataset and is only outperformed by \polars, which enables fast querying by leveraging mask operations~\cite{polars_github}, on the smaller ones.
% \footnote{\url{https://github.com/pola-rs/polars/blob/py-1.5.0/py-polars/polars/dataframe/frame.py\#L4400-L4554}}
Finally, \pandas generally maintains overall good performance for \emph{getcols} and \emph{dtypes}, where \pyspark is much slower due to the overhead associated to its inherently distributed architecture, optimized for complex tasks rather than simple ones.

\subsubsection*{Data Transformation (DT)} 

In this stage, \cudf and \polars emerge as the best performers overall, highlighting the benefits of GPU acceleration and \rust code optimizations, respectively.
In particular, \cudf outperforms \polars on \loan (Figure~\ref{fig:loan_dt}) and \patrol (Figure~\ref{fig:patrol_dt}), while the opposite is true on \athlete (Figure~\ref{fig:athlete_dt}).
As depicted in Figure~\ref{fig:speedup}, \cudf always achieves the best results for \emph{catenc}, while for \emph{onehot} and \emph{group} it is the best performer on all datasets but \athlete, where \polars outperforms the alternatives.
\sparksql shows excellent results on \patrol (Figure~\ref{fig:patrol_dt}): despite not excelling for single preparators (Figure~\ref{fig:patrol_core}), lazy evaluation provides remarkable benefits.
Moreover, it clearly outperforms \sparkpd, which results as the worst performer on several preparators, mainly due to the latency introduced by translating \pandas operations into \spark execution plans.
% Grouping on \athlete mostly involves \emph{count} aggregations, for which \polars is very efficient due to its row encoding and multithreaded counting.
% GPU acceleration provides instead a significant boost for more complex operations like \emph{mean} and \emph{sum}.}

% \textcolor{green}{As depicted in Figure~\ref{fig:speedup}, the two libraries reflect the same ranking for \emph{group}.
% This result depends on the aggregation functions used in the pipelines: while \polars is very efficient for \emph{count} due to its row encoding and multithreaded counting, GPU acceleration provides a significant boost for more complex operations like \emph{mean} and \emph{sum}.
% Further, \cudf always shows the best performance for \emph{catenc} and even on \emph{onehot}, except on \athlete.}

\vaex optimizations lead to outstanding performance on column-wise preparators (e.g., \emph{calccol} and \emph{drop}), but also to drastic drops moving to operations such as \emph{group}, \emph{join}, or \emph{pivot}.
\modin, designed for resource-intensive applications, improves its performance as datasets scale up, with \modinr being the best performer for \emph{pivot} on \taxi.
\datatable registers remarkable performance on that preparator too, despite not supporting it natively~\cite{datatable_pandas_comparison}, but particularly excels for \emph{cast}, due to direct memory manipulation and in-place casting~\cite{datatable_cast}.
On the other hand, libraries relying on the \arrow columnar format (e.g., \polars), with its safety check and abstraction layers, may experience additional overhead~\cite{polars_cast}.

\subsubsection*{Data Cleaning (DC)}

When it comes to DC, \polars achieves the best results on \athlete (Figure~\ref{fig:athlete_dc}), outperforming \cudf (which performs better for all preparators if considered separately, as depicted in Figure~\ref{fig:athlete_core}) thanks to lazy evaluation.
\cudf excels on \loan (Figure~\ref{fig:loan_dc}) thanks to its performance with \emph{dedup}, which has by far the highest runtime among the preparators, thus deserving an in-depth analysis in the following.
\cudf implements \emph{dedup} natively, by using factorization to identify duplicates~\cite{pandas_dedup}.
Differently, \polars tracks and manages the presence or absence of unique values using bitmasking~\cite{polars_github}, while \sparksql uses grouping and aggregations to find duplicate rows~\cite{pyspark_code}.
Finally, \vaex and \datatable lack an official implementation for \emph{dedup}.

Results change when moving to larger datasets (Figures~\ref{fig:patrol_dc} and~\ref{fig:taxi_dc}), with \vaex achieving the best results both on \patrol (followed closely by \cudf and \modinr) and on \taxi (with remarkable performance by \polars and both \modin versions).
On \patrol, \vaex is the best performer on both preparators (Figure~\ref{fig:patrol_core}): \emph{dropna}, where it exploits its efficient filtering mechanisms, and \emph{chdate}, for which it relies on \numpy operations (faster compared to more complex operations found in other libraries) to enhance conversion and manipulation of columns~\cite{vaex_github}.

\begin{figure*}[!ht]

\begin{subfigure}{\textwidth}
         \centering
         \hspace*{0.5cm}\includegraphics[width=0.98\linewidth]{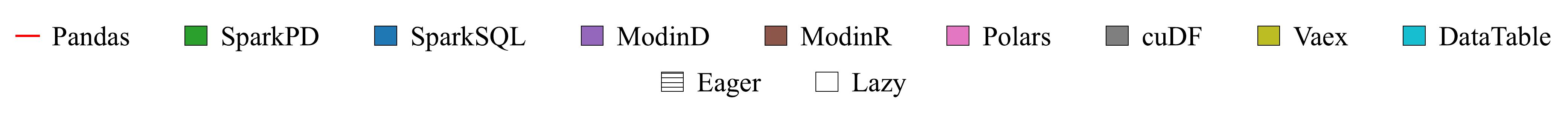}
\end{subfigure}
\begin{subfigure}{\textwidth}
    \begin{subfigure}[b]{0.245\textwidth}
         \centering
         \includegraphics[width=\textwidth]{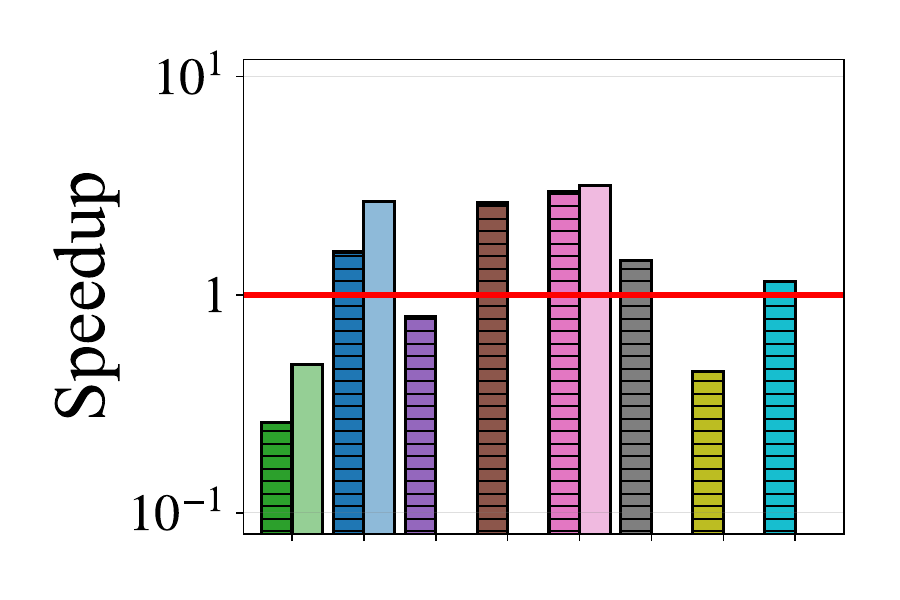}
         \vspace*{-0.6cm}
         \caption{Athlete}
         \label{fig:athlete_eager_lazy}
    \end{subfigure}
    \hfill
        \begin{subfigure}[b]{0.245\textwidth}
         \centering
         \includegraphics[width=\textwidth]{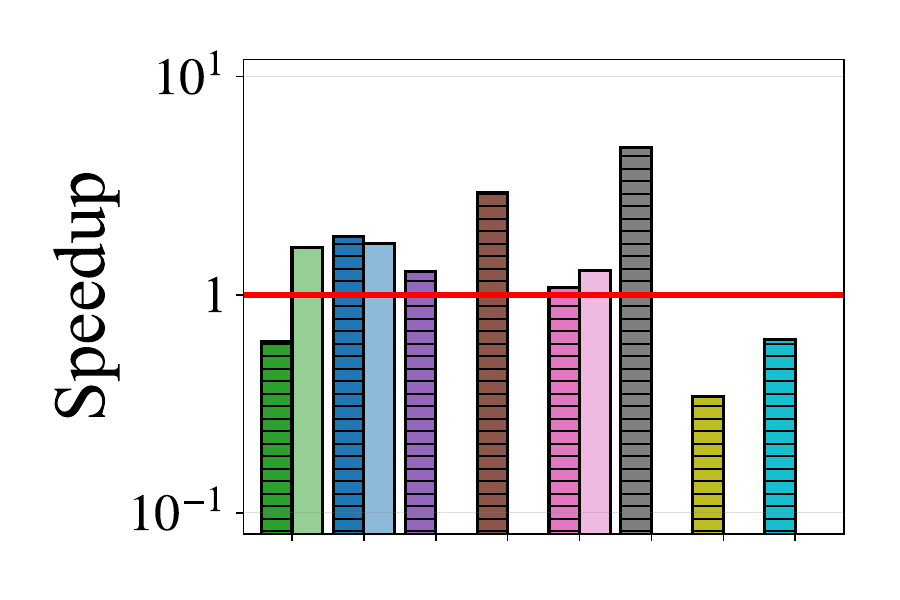}
         \vspace*{-0.6cm}
         \caption{Loan}
         \label{fig:loan_eager_lazy}
    \end{subfigure}
    \hfill
        \begin{subfigure}[b]{0.245\textwidth}
         \centering
         \includegraphics[width=\textwidth]{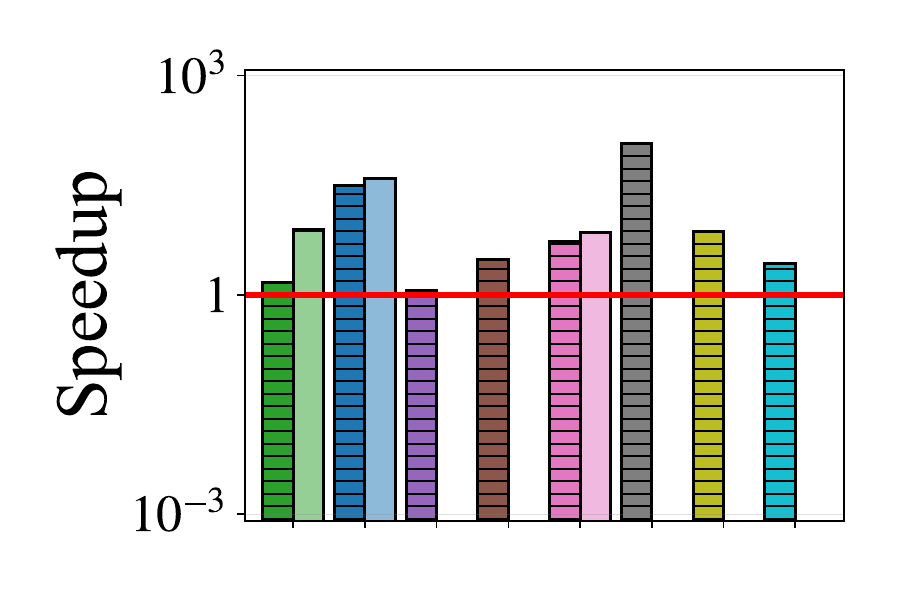}
         \vspace*{-0.61cm}
         %\hspace{0.6cm}
         \caption{Patrol}
         \label{fig:patrol_eager_lazy}
    \end{subfigure}
    \hfill
        \begin{subfigure}[b]{0.245\textwidth}
         \centering
         \includegraphics[width=\textwidth]{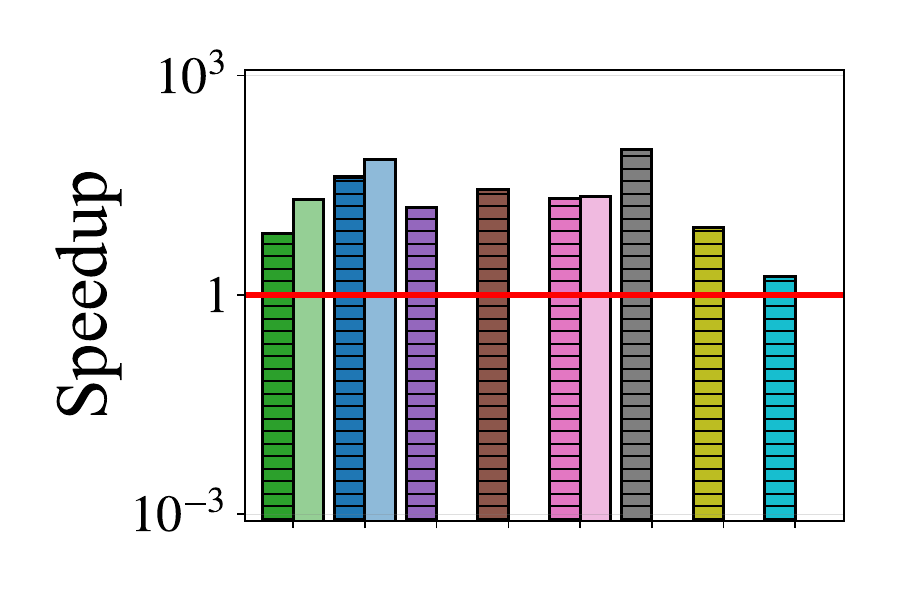}
         \vspace*{-0.6cm}
         \caption{Taxi}
         \label{fig:taxi_eager_lazy}
    \end{subfigure}
\end{subfigure}
%\vspace*{-0.5cm}
\caption{Average speedup over Pandas for the execution of the entire data preparation pipelines.}
\label{fig:eager_lazy}
\end{figure*}
% -------------------------------------------------- %

\subsubsection*{Input/Output (I/O)}

Loading data represents a fundamental step for the success of any data-driven process.
In particular, the \csv format is considered as the \emph{de facto} standard for reading and writing raw data, and \pandas is recognized as an efficient and effective tool for its ingestion~\cite{vitagliano2023pollock@pvldb}.
Nowadays, the diffusion of \arrow is pushing towards an increasing adoption of the \parquet format~\cite{parquet}, known for its efficient data compression.
% \footnote{\url{https://parquet.apache.org}}
In particular, developers often load \parquet data into memory and convert it to \arrow~\cite{liu2023analytical_open_formats@pvldb}, recognized as an ideal in-memory transport layer for such data~\cite{arrow_parquet}.
% \footnote{\url{https://arrow.apache.org/docs/python/parquet.html}}

Figures~\ref{fig:read_csv_parq} and~\ref{fig:write_csv_parq} show performance in reading and writing both file formats, \csv and \parquet.
When reading small datasets like \athlete, \parquet and \csv often exhibit similar performance, with exceptions such as \polars or \sparksql (Figure~\ref{fig:athlete_read}).
In some cases (e.g., \vaex or \cudf), the use of \parquet can even lead to worse performance, due to the need for decompressing the data.
When the size increases, the use of \parquet generally produces better results; for instance, reading \parquet files with \polars is over 100 times faster than reading \csv files (Figure~\ref{fig:taxi_read}).

Overall, \cudf and \vaex appear to be the best performers in reading \csv files.
\vaex is particularly fast thanks to chunked reading, conversion to optimized formats (e.g., \textsf{HDF5}), and efficient use of memory~\cite{vaex_io}.
\cudf, relying on GPU acceleration and efficient memory management, maintains similar performance for both file formats.
\datatable, which maps the file into memory and navigates through it using pointers, performs good in reading \csv files, but it does not support \parquet.
\polars and \vaex excel in reading \parquet files thanks to their capability to load them directly into memory using \arrow.
%\polars is the best performer in reading \parquet files due to its capability to load them directly into memory using \arrow.

% {\color{blue}
% \cudf followed by \vaex are the best performers in reading \csv files.
% \cudf is one of the fastest library thanks to GPU accelleration and efficent memory management, mantain the same performance for both the file formats.
% While \vaex uses chunked reading, conversion to optimized formats (e.g., \emph{HDF5})\footnote{\url{https://vaex.readthedocs.io/en/latest/_modules/vaex.html#from_csv}}, and efficient use of memory, are key reasons why is particularly fast at reading \csv and \parquet files.
% }

Write operations exhibit a wide range of performance.
Overall, \polars and \cudf are the top performers and \parquet performs better than \csv across all datasets as it integrates efficient compression and encoding approaches, while \csv is plain text and uncompressed~\cite{liu2023analytical_open_formats@pvldb}.
Nevertheless, the overhead introduced by compression and encoding operations can penalize \parquet in some scenarios, as highlighted by \cudf on small datasets (e.g., \athlete and \loan).
Further, \polars presents a reported issue in writing \parquet files, which determines slower performance and larger file size~\cite{polars_github}.
Finally, \modin (with both engines) demonstrates the best performance on \parquet, particularly for large datasets, by leveraging parallel processing across multiple cores~\cite{modin_documentation}.

% Write operations exhibit a wide range of performance, mainly due to the need of compressing data before saving.
% \parquet files come with additional metadata and encoding, which may introduce some overhead for the writing process. 
% Nevertheless, \parquet performs better than \csv across all datasets as it integrates various efficient compression and encoding approaches, while \csv is plain text and uncompressed~\cite{liu2023analytical_open_formats@pvldb}.
% {\color{blue}
% However, for \cudf on small datasets (e.g., \athlete and \loan), \parquet can be less efficient than \csv due to the overhead introduced by its compression and encoding mechanisms. 
% Additionally, \polars is the only library where writing \parquet consistently performs worse than \csv across all datasets. This issue stems from a known bug in \polars \parquet write operation~\footnote{\url{https://github.com/pola-rs/polars/pull/14818}}. The problem involves \polars writing more row groups by default compared to \pyarrow, leading to slower performance and larger file sizes.
% }

% For instance, while on \patrol the \parquet format brings significant advantages to \pandas (outperforming \csv by a factor of 27) thanks to its efficient dataframe serialization, \modin and \cudf are much faster in writing \csv files (Figure~\ref{fig:patrol_write}).
% Nevertheless, \csv format causes \cudf to run out of memory on the largest dataset (Figure~\ref{fig:taxi_write}), while it succeeds in writing \parquet files.

\subsection{Impact of Lazy Evaluation}
\label{sec:evalstrategies}
\smallskip
\begin{mdframed}
    \textit{Summary}---\cudf generally outperforms other libraries when running the entire pipeline, while \sparksql and \modinr achieve remarkable results on all datasets.
    Lazy evaluation often brings relevant benefits, improving performance by 20\% on average (up to almost 80\%) over its eager counterpart.
\end{mdframed}
\smallskip
Figure~\ref{fig:eager_lazy} reports the average speedup over \pandas achieved by its alternatives for the execution of the entire data preparation pipelines on each dataset.
For libraries supporting lazy evaluation (i.e., \sparkpd, \sparksql, and \polars), we also point out the difference compared to eager evaluation.

Considering the overall performance on the entire pipeline, \cudf generally outperforms other libraries by effectively exploiting the power of GPU, with the only exception of \athlete (Figure~\ref{fig:athlete_eager_lazy}),  where \polars is the best performer.
\pyspark (except for \sparkpd on \athlete), \modinr, and \polars show very solid performance overall, as well as \vaex on the largest datasets.
\pandas performs relatively well on the smallest datasets, but it is outperformed by all alternatives on larger ones, due to its eager approach causing high memory consumption~\cite{sinthong2021polyframe@pvldb}.

\polars generally achieves good results with both kinds of evaluations, as its eager API is highly optimized and in many cases it internally relies on its lazy counterpart before immediately collecting the result.
Lazy evaluation leverages techniques such as streaming processing, early filtering, and projection pushdown, achieving a performance improvement of up to 25\% on \patrol (Figure~\ref{fig:patrol_eager_lazy}).
On one hand, it can be noticed how long logical plans can limit the optimizations introduced by lazy evaluation. Yet, the execution plan overhead of \sparksql does not yield substantial improvements for either smaller or larger datasets, with an improvement of 40\% on \taxi and almost no changes in performance on \loan.
Libraries like \sparkpd, leverage to \pandas-specific optimizations (e.g., vectorized operations and efficient memory management) that translate for instance into a 80\% performance improvement on \patrol (Figure~\ref{fig:patrol_eager_lazy}).

\subsection{Scalability}
\label{sec:scalability}
\smallskip
\begin{mdframed}
    \textit{Summary}---\sparksql is the best library for scaling to large datasets on a single machine.
    Indeed, for both \patrol and \taxi, it is the only one able to execute the entire pipelines with the laptop configuration. 
    \polars follows closely, but requires a lot of resources in terms of RAM, exhibiting the worst scalability on machines with limited resources.
\end{mdframed}
\smallskip
In this section, we assess how dataframe libraries scale by varying the size of the dataset and the configuration of the underlying machine.
In particular, we take into account three different machine configurations, whose CPU and RAM are incrementally increased as described in Table~\ref{tab:machineconfiguration} (which also reports the specifications of the \dask and \ray engines used by \modin) to reproduce typical configurations for laptops, workstations, and servers.

\begin{table}[!t]
\small
\centering
\caption{Specifications of each machine configuration.}
%\vspace{-0.1cm}
\label{tab:machineconfiguration}
\begin{adjustbox}{max width=\columnwidth}
\begin{tabular}{r|ccc|}
\cline{2-4}
                                & \textbf{Laptop} & \textbf{Workstation} & \textbf{Server} \\ \hhline{-===}
\multicolumn{1}{|r|}{\# CPUs}   & 8               & 16                   & 24              \\ \hline
\multicolumn{1}{|r|}{\# RAM (GB)} & 16              & 64                   & 128             \\ \hline
\multicolumn{1}{|r|}{Dask (workers - threads)} & 4-8              & 4-16                   & 6-24             \\ \hline
\multicolumn{1}{|r|}{Ray (workers)} & 8              & 16                   & 24             \\ \hline
\end{tabular}
\end{adjustbox}
\end{table}

For this experiment, we selected the most computationally expensive pipeline among the three evaluated, which is the first one\footnote{On average, this pipeline is approximately three times more computationally expensive than the others.}.
Since \cudf scales according to GPU memory, it is not included in this experiment.
If the dataset fits in the GPU, \cudf is able to efficiently handle it and complete the pipeline.
However, this dependency does not make it a good candidate for dealing with very large datasets on machines with limited GPU resources.

\begin{figure}[!t]
    \centering
    \begin{subfigure}{\columnwidth}
             \centering
             \hspace*{0.4cm}\includegraphics[width=0.97\linewidth]{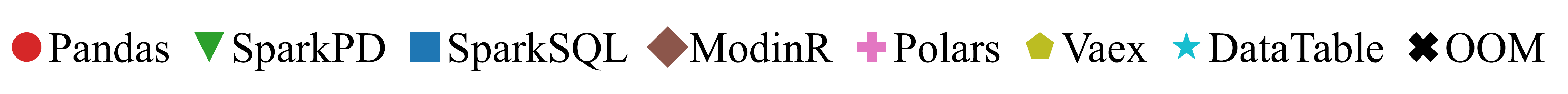}
    \end{subfigure}
    \begin{subfigure}{\columnwidth}
        \begin{subfigure}[b]{0.3\textwidth}
             \centering
             \includegraphics[width=1.11\textwidth]{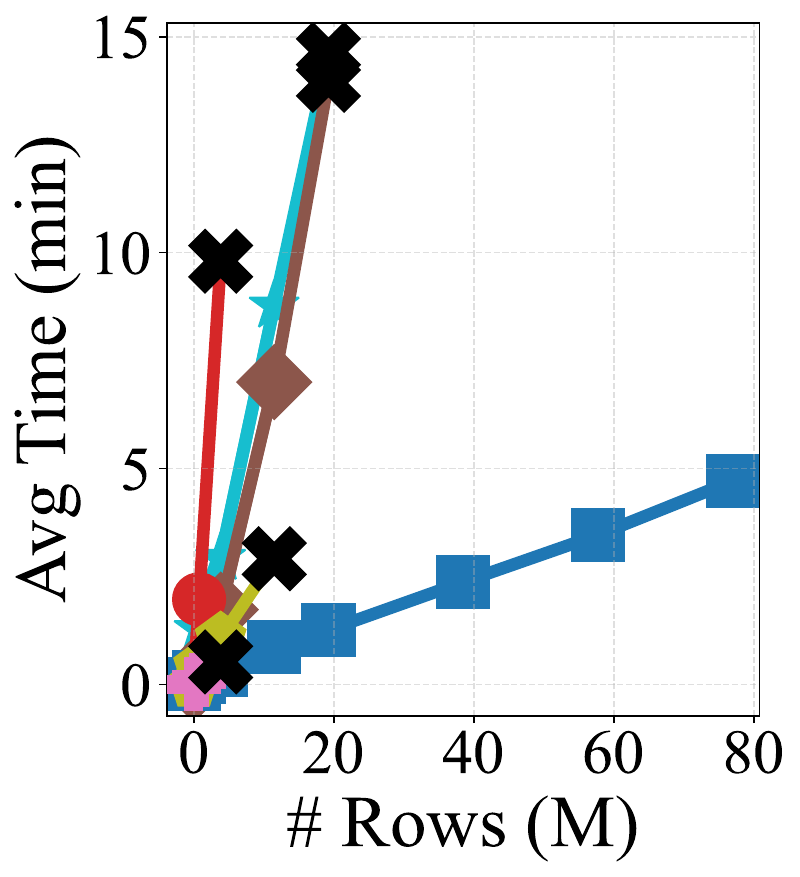}
             \vspace*{-0.4cm}
             \caption{Laptop}
             \label{fig:8_16}
        \end{subfigure}
        \hspace{-0.4cm}
        \hfill
            \begin{subfigure}[b]{0.3\textwidth}
             \centering
             \includegraphics[width=\textwidth]{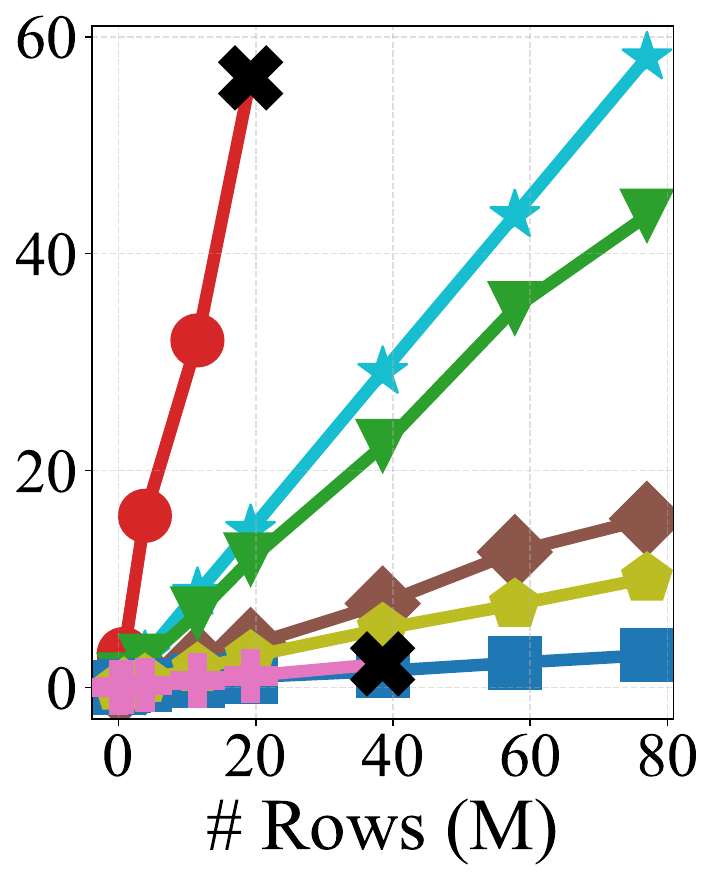}
             \vspace*{-0.4cm}
             \caption{Workstation}
             \label{fig:16_64}
        \end{subfigure}
        \hspace{-0.7cm}
        \hfill
            \begin{subfigure}[b]{0.3\textwidth}
             \centering
             \includegraphics[width=1.04\textwidth]{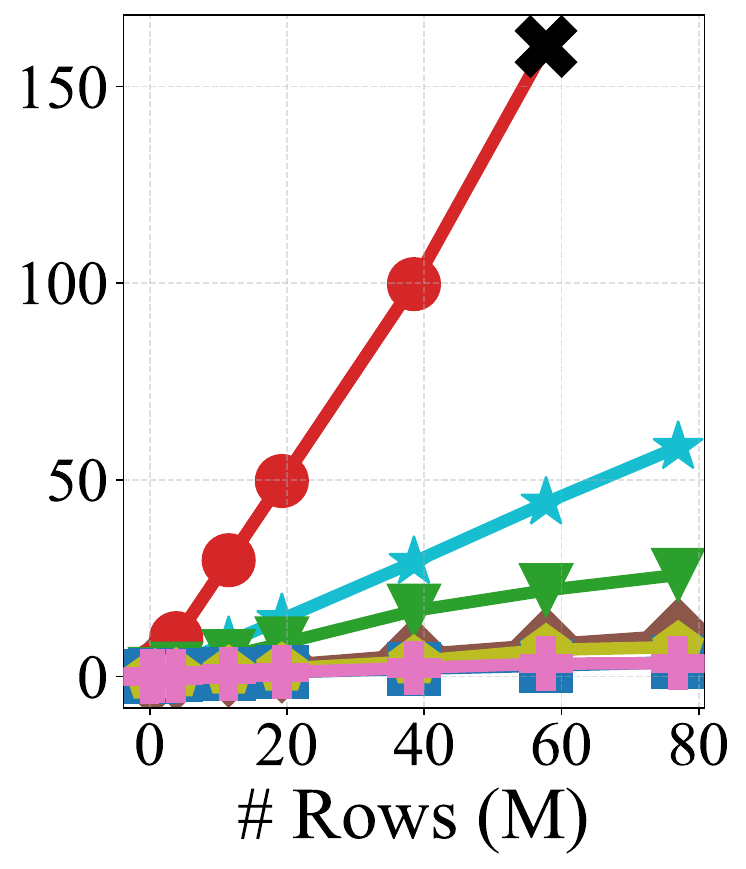}
             \vspace*{-0.4cm}
            \caption{Server}
             \label{fig:24_128}
        \end{subfigure}
    
    \end{subfigure}
    %\vspace{-0.6cm}
    \caption{Average runtime for running the entire pipeline on incremental samples of \taxi.}
    \label{fig:scalability}
\end{figure}

Figure~\ref{fig:scalability} shows the performance of the libraries for running the entire pipeline on incremental samples of \taxi.
\sparksql clearly stands out as the best performer, being the only library capable of handling  the full pipeline execution on laptop configuration (Figure~\ref{fig:8_16}). This capability is attributed to its combination of lazy evaluation and disk spillover mechanisms~\cite{pyspark_diskspill}---optimizing execution plans and offloading data to disk when memory limits are reached during computations.
%use of disk spillover~\cite{pyspark_diskspill}, which ensures continuous operation when memory limits are reached during computations.
\sparkpd suffers instead from the overhead due to internal conversions between \pandas and \spark dataframes and increased \textsf{JVM} memory usage, causing out-of-memory  (OOM) errors on the laptop configuration.

While \modin (both \dask and \ray) supports operations on datasets larger than available memory~\cite{modin_documentation}, the efficacy of this feature is constrained by two primary factors.
Firstly, it requires additional memory for managing internal operations, metadata, and intermediate results.
Secondly, certain operations (e.g., apply lambda function), necessitate in-memory data access. 
\datatable shares this limitation as well, as a results
%, where complex operations requiring execution on each row of the dataframe can lead to OOM errors, regardless of the dataset size.
these two libraries encounter OOM issues when processing datasets larger than 25\% of the full dataset.
Similarly, \vaex experiences OOM errors with datasets exceeding 15\% of the full dataset size. 
This is due to the fact that in \vaex the output of some operations (e.g., groupby) is held entirely in memory, potentially exceeding available RAM~\cite{vaex_github}.
%~\footnote{\url{https://github.com/vaexio/vaex/issues/2276}}

Despite \polars excellent performance in other experiments, scalability appears its weakness---it reaches OOM with 4 million and 40 million rows in Figures~\ref{fig:8_16} and ~\ref{fig:16_64}, respectively.
This limitation arises from its in-memory execution model, which requires all data to be loaded into RAM. While this approach offers speed and efficiency for smaller datasets, it becomes a bottleneck when handling larger datasets, limiting its scalability.
Finally, \pandas confirms its well-known severe limitations about scalability, being the only library not able to complete the pipeline on \taxi even on a server configuration (Figure \ref{fig:24_128}).

To provide a comprehensive overview of how libraries scale across datasets with different characteristics, Table~\ref{tab:scalability} outlines instead the minimum configuration required by each library to successfully run the full pipeline on progressively larger samples of \patrol and \taxi.

\sparksql is the only library able to run the pipelines with the laptop configuration on the entire datasets---thanks to lazy evaluation and disk spillover.
%Besides that, the libraries show the same behavior across the two datasets.
The second best in scalability is \datatable, which excels on \patrol requiring minimal resources through efficient memory mapping on disk.
%Indeed, {\color{blue}\datatable} is one of the libraries that consumes the least amount of memory on \patrol {\color{blue}  is the only one running 50\% of the dataset in the laptop configuration};} 
%Further, it does not require to split the dataframe into multiple pieces, which can lead to errors and inefficiencies due to the complexities of communication and information exchange among partitions. 

%{\color{blue}Libraries like \modin and \spark—even when running on a single machine—split data into multiple partitions to parallelize tasks, introducing potential challenges related to data synchronization and overhead.}
%further, it does not require to split the dataframe into multiple pieces, {\color{red}which can lead to errors due to communication and information exchange.}

% However, it shows poor performance on \taxi, where it manages to complete the pipeline only on a workstation configuration (Figure~\ref{fig:16_64})
% \textcolor{red}{Note that \modin exhibits the same performance with both engines.}

\modinr demonstrates better scalability across all machine configurations compared to its \dask counterpart. The \dask engine uses a centralized scheduler to distribute data across multiple cores, which results in higher memory consumption and leads to OOM issues more easily~\cite{dask_documentation}. In contrast, \ray can scale far beyond Dask due to its distributed task scheduling scheme, specifically employing a distributed bottom-up scheduling approach~\cite{ray_documentation}. This optimizes resource utilization and minimizes overhead by initiating tasks at the lowest level of computation and aggregating results upwards. Notably, \modin was originally built to work on top of \ray, making this integration more mature and optimized~\cite{modin_documentation}.

\begin{figure*}[!t]
\begin{subfigure}{\textwidth}
         \centering
         \includegraphics[width=\linewidth]{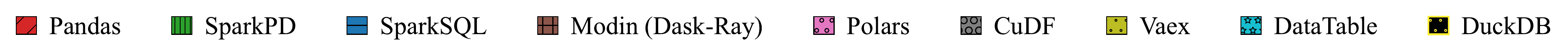}
\end{subfigure}
\begin{subfigure}{\textwidth}
         \centering
         \includegraphics[width=\linewidth]{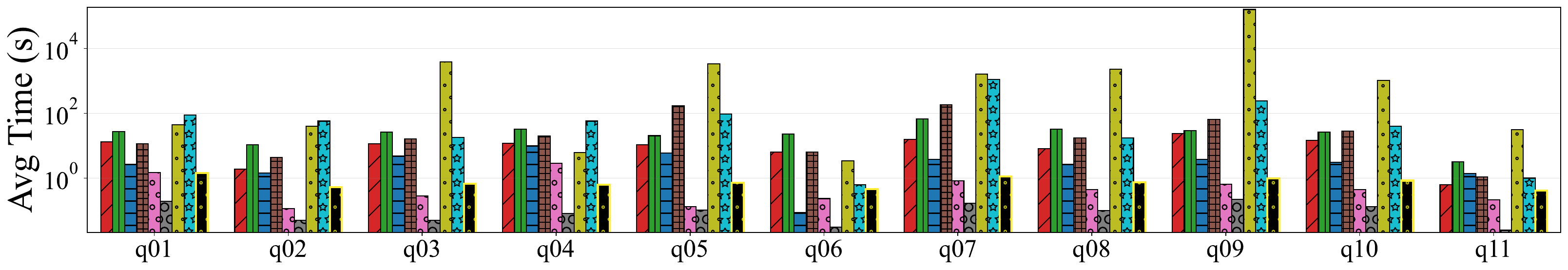}
         \vspace{-0.5cm}
         \caption{Queries 1 to 11.}
         \label{fig:tpch_1_11.pdf}
         \vspace{10pt}
\end{subfigure}
\hfill
\begin{subfigure}{\textwidth}
         \centering
          \includegraphics[width=\linewidth]{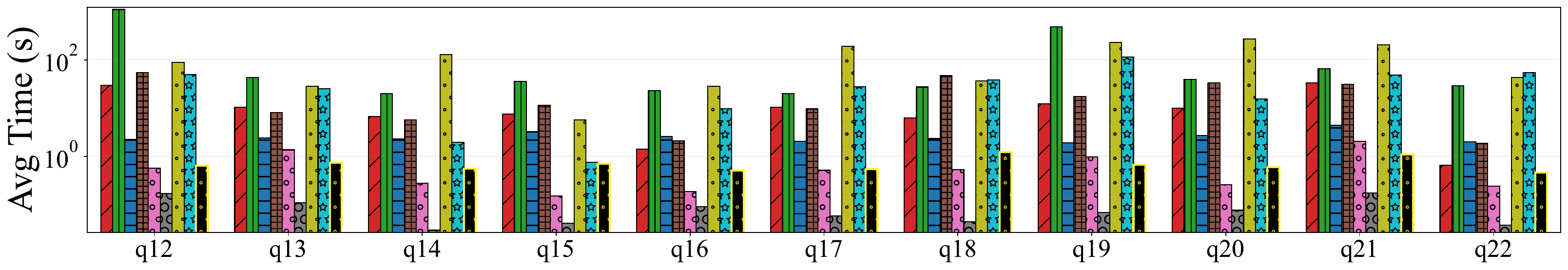}
          \vspace{-0.5cm}
         \caption{Queries 12 to 22.}
         \label{fig:tpch_12_22}
\end{subfigure}
%\vspace{-0.6cm}
\caption{Performance of the dataframe libraries on the TPC-H 10GB queries (lower is better).}
\label{fig:tpch}
\end{figure*}

\subsection{Performance on the TPC-H Queries}
\label{sec:tpch}
\smallskip
\begin{mdframed}
    \textit{Summary}---\cudf consistently achieves the best performance across all queries, while \polars significantly outperforms other CPU-only dataframe libraries.
    % --thanks to 40GB of GPU memory, on TPC-H 100GB it will go OOM.
    % Looking at cpu-only library \polars show the best results.
\end{mdframed}
%\footnote{\url{https://www.tpc.org/tpch/default5.asp}}
\medskip
TPC-H~\cite{tpch_benchmark} is a decision support benchmark consisting of a suite of business-oriented ad-hoc queries and concurrent data modifications~\cite{poss2000tpc_benchmark@sigmod_record}.
It is widely adopted to compare end-to-end database systems~\cite{dreseler2020quantifying}, allowing to comprehensively evaluate their performance and efficiency through the execution of complex queries on large volumes of data.
Consistently with previous related work~\cite{polars_tpch_benchmark} exploiting TPC-H queries to assess the performance of a subset of dataframe libraries (see Section~\ref{sec:related_work} for more details), we also adopt this benchmark to provide further support to the outcomes of the comparative analysis presented above.

% \footnote{\url{https://pola.rs/posts/benchmarks/}}
%\footnote{\url{https://github.com/Bodo-inc/Bodo-examples}}

In particular, we take as a reference a publicly available translation of the TPC-H queries into \pandas API operations~\cite{bodo_queries}, replicating it for the other libraries, then use \bento for their seamless execution.
Figure \ref{fig:tpch} illustrates the performance of each library on the 22 queries of the TPC-H 10GB benchmark.
We select a scale factor of 10, because it represents the largest dataset that can be processed by the 40 GB RAM of our GPU, which therefore would not be able to handle a scale factor of 100.
Moreover, we included \duckdb~\cite{raasveldt2019duckdb_demo@sigmod}, being a popular in-process SQL analytical database management system.
%We chose \duckdb for this comparison due to its popularity among data scientists and its ability to efficiently process in-memory data, making it conceptually closer to dataframe libraries than traditional \sql databases. 
\duckdb can execute parallel queries directly on \pandas DataFrames, Parquet/CSV files, or Arrow tables without a separate import step, and can write results back to these formats. 
This interoperability allows seamless integration with other data science libraries. We executed the TPC-H queries on \duckdb using the provided implementation in the Polars benchmark~\cite{polars_tpch_benchmark}.
Notice that \duckdb does not provide \pandas API and only supports \sql. 
Thus, by expressing the pipelines as \sql queries, it can take advantage of all the well-known query optimization and execution strategies typical of relational database management systems.
For this reason, we do not compare it with the other libraries throughout the paper, but only report its performance with TPC-H to provide a valuable reference point w.r.t. OLAP database management systems.

%It is important to note that while \duckdb results offer an interesting comparison, they are not directly comparable to the dataframe-based approaches due to fundamental differences in query execution and optimization strategies. 
%We include these results to provide a broader perspective on query performance across different data processing paradigms.
Overall, \cudf consistently achieves the best performance by leveraging GPU power.
% However, note that it would not be able to successfully run the benchmark if the scale factor increases to 100, because it is strongly dependent on GPU memory (in our case, 40 GB).
Among the CPU-only libraries, \polars is instead the clear winner. 
This confirms what is observed in the data preparation pipelines, for EDA and DT.
\sparksql~ outperforms \pandas on several queries, registering notable results, especially on q06, where it is even faster than \polars (note that this query selects line items shipped within a year interval, reducing the dataset by 85\%). 
By leveraging lazy evaluation, \sparksql significantly reduces the amount of data processed early in the query execution---this is one of the key reasons why it improves performance in data preparation pipelines by approximately 60\%.
% This is because q06 selects line items shipped within a twelve-month frame, reducing the dataset by 85\%. \sparksql advanced optimization techniques, efficiently handle these filtering and aggregation operations.
\pandas is never the worst performer, but on most queries, it is among the slowest.
\modin (which registers the same results with both engines, \dask and \ray) shows diverse performance instead, strongly dependent on the query, confirming the inconsistent performance observed across the different data preparation stages.
% \footnote{We do not consider \modind since the results on TPC-H are the same of its \ray counterpart.} 

As mentioned above, \sparkpd suffers from the additional latency due to the translation of \pandas operations into \spark execution plans, making it one of the worst performers.
\datatable is one of the worst performers too, since it is slow in grouping operations (as also testified by the execution of \emph{group} preparator in DT stage and their own benchmark~\cite{h2o_benchmark}) and its API only supports joins on columns with unique values, requiring therefore to switch to the default \pandas API (e.g., q09).
% \footnote{\url{https://h2oai.github.io/db-benchmark/}}
\vaex is by far the worst performer overall, as it is significantly slow in grouping operations (as also pointed out in the previous sections) and lacks support for multi-column joins.

\newcommand{\sv}{\selectfont{\cellcolor[HTML]{f9844a}{\textsf{}\xspace}}}
\newcommand{\pc}{\selectfont{\cellcolor[HTML]{90BE6D}{\textsf{}\xspace}}}
\newcommand{\ws}{\selectfont{\cellcolor[HTML]{FEDC97}{\textsf{}\xspace}}}

\begin{table}[!t]
    \caption{Minimum machine configuration for running the entire pipeline on incremental dataset samples.}
    \label{tab:scalability}
    \centering
    \resizebox{0.48\textwidth}{!}{%
        \huge
        \begin{tabular}{r|ccccccc|@{\hspace{5pt}}|ccccccc|}
    \cline{2-15}
     &
      \multicolumn{7}{c|@{\hspace{5pt}}|}{Patrol} &
      \multicolumn{7}{c|}{Taxi} \\\hhline{-==============}
    \multicolumn{1}{|r|}{\textbf{\% Sample}} &
      \multicolumn{1}{c|}{\textbf{1\%}} &
      \multicolumn{1}{c|}{\textbf{5\%}} &
      \multicolumn{1}{c|}{\textbf{15\%}} &
      \multicolumn{1}{c|}{\textbf{25\%}} &
      \multicolumn{1}{c|}{\textbf{50\%}} &
      \multicolumn{1}{c|}{\textbf{75\%}} &
      \textbf{100\%} &
      \multicolumn{1}{c|}{\textbf{1\%}} &
      \multicolumn{1}{c|}{\textbf{5\%}} &
      \multicolumn{1}{c|}{\textbf{15\%}} &
      \multicolumn{1}{c|}{\textbf{25\%}} &
      \multicolumn{1}{c|}{\textbf{50\%}} &
      \multicolumn{1}{c|}{\textbf{75\%}} &
      \textbf{100\%}\\ \cline{1-15}
      \multicolumn{1}{|r|}{\textbf{\# Rows (M)}} &
      \multicolumn{1}{c|}{\textbf{0.2}} &
      \multicolumn{1}{c|}{\textbf{1.2}} &
      \multicolumn{1}{c|}{\textbf{4.5}} &
      \multicolumn{1}{c|}{\textbf{6.7}} &
      \multicolumn{1}{c|}{\textbf{13.5}} &
      \multicolumn{1}{c|}{\textbf{20.2}} &
      \textbf{27} &
      \multicolumn{1}{c|}{\textbf{0.7}} &
      \multicolumn{1}{c|}{\textbf{3.8}} &
      \multicolumn{1}{c|}{\textbf{11.5}} &
      \multicolumn{1}{c|}{\textbf{19.2}} &
      \multicolumn{1}{c|}{\textbf{38.8}} &
      \multicolumn{1}{c|}{\textbf{57.7}} &
      \textbf{77}\\ \hline \hline
    \multicolumn{1}{|r|}{Pandas} &
      \multicolumn{1}{c|}{\pc} &
      \multicolumn{1}{c|}{\pc} &
      \multicolumn{1}{c|}{\pc} &
      \multicolumn{1}{c|}{\pc} &
      \multicolumn{1}{c|}{\ws} &
      \multicolumn{1}{c|}{\ws} &
      \ws &
      \multicolumn{1}{c|}{\pc} &
      \multicolumn{1}{c|}{\pc} &
      \multicolumn{1}{c|}{\ws} &
      \multicolumn{1}{c|}{\ws} &
      \multicolumn{1}{c|}{\sv} &
      \multicolumn{1}{c|}{\sv} &
      \xmark\\ \hline
    \multicolumn{1}{|r|}{SparkPD} &
      \multicolumn{1}{c|}{\ws} &
      \multicolumn{1}{c|}{\ws} &
      \multicolumn{1}{c|}{\ws} &
      \multicolumn{1}{c|}{\ws} &
      \multicolumn{1}{c|}{\ws} &
      \multicolumn{1}{c|}{\ws} &
      \ws &
      \multicolumn{1}{c|}{\ws} &
      \multicolumn{1}{c|}{\ws} &
      \multicolumn{1}{c|}{\ws} &
      \multicolumn{1}{c|}{\ws} &
      \multicolumn{1}{c|}{\ws} &
      \multicolumn{1}{c|}{\ws} &
      \ws\\ \hline
    \multicolumn{1}{|r|}{SparkSQL} &
      \multicolumn{1}{c|}{\pc} &
      \multicolumn{1}{c|}{\pc} &
      \multicolumn{1}{c|}{\pc} &
      \multicolumn{1}{c|}{\pc} &
      \multicolumn{1}{c|}{\pc} &
      \multicolumn{1}{c|}{\pc} &
      \pc &
      \multicolumn{1}{c|}{\pc} &
      \multicolumn{1}{c|}{\pc} &
      \multicolumn{1}{c|}{\pc} &
      \multicolumn{1}{c|}{\pc} &
      \multicolumn{1}{c|}{\pc} &
      \multicolumn{1}{c|}{\pc} &
      \pc\\ \hline
    \multicolumn{1}{|r|}{ModinD} &
      \multicolumn{1}{c|}{\pc} &
      \multicolumn{1}{c|}{\pc} &
      \multicolumn{1}{c|}{\pc} &
      \multicolumn{1}{c|}{\ws} &
      \multicolumn{1}{c|}{\sv} &
      \multicolumn{1}{c|}{\sv} &
      \sv &
      \multicolumn{1}{c|}{\pc} &
      \multicolumn{1}{c|}{\pc} &
      \multicolumn{1}{c|}{\pc} &
      \multicolumn{1}{c|}{\ws}  & \multicolumn{1}{c|}{\sv} &
      \multicolumn{1}{c|}{\xmark} &
      \xmark\\ \hline
    \multicolumn{1}{|r|}{ModinR} &
      \multicolumn{1}{c|}{\pc} &
      \multicolumn{1}{c|}{\pc} &
      \multicolumn{1}{c|}{\pc} &
      \multicolumn{1}{c|}{\pc} &
      \multicolumn{1}{c|}{\ws} &
      \multicolumn{1}{c|}{\ws} &
      \ws &
      \multicolumn{1}{c|}{\pc} &
      \multicolumn{1}{c|}{\pc} &
      \multicolumn{1}{c|}{\pc} &
      \multicolumn{1}{c|}{\pc} &
      \multicolumn{1}{c|}{\ws} &
      \multicolumn{1}{c|}{\ws} &
      \ws\\ \hline
    \multicolumn{1}{|r|}{Polars} &
      \multicolumn{1}{c|}{\pc} &
      \multicolumn{1}{c|}{\pc} &
      \multicolumn{1}{c|}{\pc} &
      \multicolumn{1}{c|}{\ws} &
      \multicolumn{1}{c|}{\ws} &
      \multicolumn{1}{c|}{\ws} &
      \ws &
      \multicolumn{1}{c|}{\pc} &
      \multicolumn{1}{c|}{\pc} &
      \multicolumn{1}{c|}{\ws} &
      \multicolumn{1}{c|}{\ws} &
      \multicolumn{1}{c|}{\ws} &
      \multicolumn{1}{c|}{\sv} &
      \sv\\ \hline
    % \multicolumn{1}{|r|}{cuDF} &
    %   \multicolumn{1}{c|}{\pc} &
    %   \multicolumn{1}{c|}{\pc} &
    %   \multicolumn{1}{c|}{\pc} &
    %   \multicolumn{1}{c|}{\pc} &
    %   \multicolumn{1}{c|}{\pc} &
    %   \multicolumn{1}{c|}{\pc} &
    %   \pc &
    %   \multicolumn{1}{c|}{\pc} &
    %   \multicolumn{1}{c|}{\pc} &
    %   \multicolumn{1}{c|}{\pc} &
    %   \multicolumn{1}{c|}{\ws} &
    %   \multicolumn{1}{c|}{\ws} &
    %   \multicolumn{1}{c|}{\ws} &
    %   \ws\\ \hline
    \multicolumn{1}{|r|}{Vaex} &
      \multicolumn{1}{c|}{\pc} &
      \multicolumn{1}{c|}{\pc} &
      \multicolumn{1}{c|}{\pc} &
      \multicolumn{1}{c|}{\ws} &
      \multicolumn{1}{c|}{\ws} &
      \multicolumn{1}{c|}{\ws} &
      \ws &
      \multicolumn{1}{c|}{\pc} &
      \multicolumn{1}{c|}{\pc} &
      \multicolumn{1}{c|}{\pc} &
      \multicolumn{1}{c|}{\ws} &
      \multicolumn{1}{c|}{\ws} &
      \multicolumn{1}{c|}{\ws} &
      \ws\\ \hline
    \multicolumn{1}{|r|}{DataTable} &
      \multicolumn{1}{c|}{\pc} &
      \multicolumn{1}{c|}{\pc} &
      \multicolumn{1}{c|}{\pc} &
      \multicolumn{1}{c|}{\pc} &
      \multicolumn{1}{c|}{\pc} &
      \multicolumn{1}{c|}{\ws} & 
      \ws &
      \multicolumn{1}{c|}{\pc} &
      \multicolumn{1}{c|}{\pc} &
      \multicolumn{1}{c|}{\pc} &
      \multicolumn{1}{c|}{\pc} &
      \multicolumn{1}{c|}{\ws} &
      \multicolumn{1}{c|}{\ws} &
      \ws\\ \hline
      \multicolumn{15}{c}{\rule{0pt}{3ex}\colorbox[HTML]{90BE6D}{\color[HTML]{90BE6D}III} Laptop \hspace{0.25cm} \colorbox[HTML]{FEDC97}{\color[HTML]{FEDC97}III} Workstation \hspace{0.25cm} \colorbox[HTML]{F9844A}{\color[HTML]{F9844A}III} Server\hspace{0.25cm} \xmark\hspace{0.1cm}OOM}
    \end{tabular}
    }
\end{table}

%\duckdb generally underperforms compared to \cudf and \polars.
After \cudf, the second best performers are \duckdb and \polars.
They excel in different aspects of the benchmark, with neither consistently outperforming the other across all queries---they adopt two different approaches for query optimizations.

\section{Related Work}
\label{sec:related_work}
%{\color{red}Dove diciamo che non consideriamo SQL e DuckDB?.
%Io direi qualcosa del tipo che c'e' una gran letteratura su comparison di DBMS vari (e.g., TPC-H TPC-DS etc.), ma l'obiettivo nostro e' di proporre qualcosa di diverso---evaluation di data preparation pipeline con dataframe}
To the best of our knowledge, our paper presents the first rigorous and extensive experimental comparison of existing dataframe libraries on data preparation tasks.
In fact, public wisdom about such libraries is mostly scattered across several not peer-reviewed sources~\cite{schmitt2020scaling_pandas@data_revenue, alexander2023beyond_pandas@medium, pinner2023dataframes@medium}, which generally compare the performance of few libraries (e.g., \pandas vs \pyspark vs \polars) on a handful of operations.
% karlsson2023dataframes@medium (dead URL)
% polars_alternatives@polars
The major claims by such analysis are mostly confirmed by our evaluation; nevertheless, despite often providing some useful insights, they only offer a partial and very fragmented knowledge, far from a complete and detailed overview covering all libraries and operations.
In the literature, only partial comparisons have been performed among libraries, which we list in the following.
%In literature, when a paper introduces a new dataframe library or dataframe-based tool, it usually includes a comparison against some dataframe libraries.

%The \textsf{AFrame} data analysis package~\cite{sinthong2019aframe@bigdata} (later extended to \textsf{PolyFrame}~\cite{sinthong2021polyframe@pvldb}), designed to efficiently manage large-scale dataframes using SQL++ queries.

\textsf{AFrame}~\cite{sinthong2019aframe@bigdata, sinthong2021polyframe@pvldb} reports an evaluation against \pandas, \spark, and \modinr through a micro-benchmark using the synthetic Wisconsin benchmark dataset~\cite{dewitt1993wisconsin_benchmark@chapter} to assess acceleration and scalability across distributed environments.
Petersohn et al.~\cite{petersohn2021modin@pvldb} evaluate the benefits of \modin by comparing against \dask, \koalas, and \pandas.
\textsf{Grizzly}~\cite{DBLP:conf/btw/KlabeH21} evaluates against \modin and \pandas, while translating dataframe pipelines to SQL.
Shanbhag et al.~\cite{shanbhag2023dataframe_energy_consumption@msr} compare \pandas, \vaex and \dask on a set of single operations to gain insights into their energy consumption.
%However, these comparisons share significant limitations, since data preparation is not their main focus and only a relatively small set of libraries and operations is considered.

Dataframe libraries have been compared also in data science benchmarks.
For instance, \polars~\cite{polars_tpch_benchmark} compares against \pandas, \pyspark, \dask, \modin, and \duckdb on TPC-H~\cite{poss2000tpc_benchmark@sigmod_record}.
% \footnote{\url{https://pola.rs/posts/benchmarks/}}
The outcome of their evaluation is consistent with our findings (we adopt the same scale factor for TPC-H), but their evaluation only covers the first 7 queries out of 22.
Similarly, \datatable~\cite{h2o_benchmark} uses a set of queries to asses the scalability of popular database and dataframe-like systems only limited to group by and join operations.
\textsf{Sanzu}~\cite{watson2017sanzu@bigdata} proposes a benchmark designed to evaluate five popular data science frameworks and systems (\textsf{R}, \textsf{Anaconda}~\cite{anaconda}, \dask, \textsf{PostgreSQL}~\cite{postgresql}, and \pyspark) on data processing and analysis tasks.
It comprises both a micro-benchmark for testing basic operations in isolation and a macro-benchmark for evaluating series of operations representing concrete data science scenarios on real-world datasets.
Similarly, \textsf{FuzzyData}~\cite{rehman2022fuzzydata@testdb} is a workflow generation system that allows to compare dataframe-based APIs on workflows composed of a small subset of operations.
%However, some of these frameworks focus primarily on exploratory data analysis (EDA).

% \footnote{\url{https://h2oai.github.io/db-benchmark}}
%Indeed, due to their different focus, such relational benchmarking systems notoriously offer very limited insights into real-world data preparation workflows executed on dataframes~\cite{rehman2022fuzzydata@testdb}.
%{\color{red}To be Fixed: Note that for producing a consistent a analysis we do not take into account tools that are not based on the dataframe data structure, as the aforementioned \textsf{DuckDB} or similar \textsf{SQL}-based solutions.}

%{\color{red}Io metterei solo una nota da qualche parte sul fatto che non consideriamo SQL/relational systems.}

%Other works have expanded the evaluation scope of dataframe libraries beyond individual operations to encompass broader data processing workflows. 
%While single-operation assessments offer valuable insights, they often lack the holistic perspective needed for real-world scenarios. 
%To bridge this gap, several frameworks have emerged. 
Other works have explored the problem of optimizing/re-writing dataframe pipelines, reporting comparative evaluations of popular libraries.
For instance, \textsf{PyFroid}~\cite{DBLP:conf/edbt/EmaniFC24} translates from the \pandas API to the \textsf{DuckDB} API to efficiently scale \pandas workloads on a commodity workstation, using top-voted Kaggle notebooks for the evaluation.
\textsf{Dataprep}~\cite{DBLP:conf/sigmod/PengWLBYXCRW21} is a task-centric EDA tool that can use different dataframe libraries as back-end engines, including \modin, \pyspark, and \dask. 
\textsf{Dias}~\cite{DBLP:journals/pacmmod/BaziotisKM24} is a system for dynamically rewriting \pandas code to optimize performance.
All of these studies offer valuable insights into dataframe library performance across some real-world scenarios, but they do not provide a comprehensive evaluation on data preparation tasks with all dataframe libraries considered here.

\section{Key Takeaways}
\label{sec:key_takeaways}

Our evaluation confirmed the well-known limitations of \pandas.
However, despite the abundance of dataframe libraries designed to overtake it, there is no silver bullet to perform the preparation of tabular data with dataframes on a single machine.

The relative performance of the libraries varies significantly among the variegate datasets and stages of the data preparation pipelines that we considered.
None of the tools emerges as the clear winner for all considered scenarios.
Nevertheless, certain characteristics of the dataset and the pipeline can help narrow the user's choice of the tool. Below, we list three questions to assist in this selection.

%The performance of each library varies significantly depending on multiple factors, making it challenging to draw universally applicable conclusions.

%The right choice usually depends on many factors, such as the size and the features of the dataset at hand, the configuration of the underlying machine, and the kind of operations to carry out.
%{\color{green}While we can provide some general guidelines to help data practitioners decide which library to use, these recommendations are not definitive and may not apply to all scenarios.}
%We can help data practitioners to decide which library to use by answering the following questions.

% -------------------------------------------------- %

\subsubsection*{Does the dataset fit in the GPU memory?}

If a GPU with enough memory to fit the data is available, \cudf appears to consistently yield the best overall performance and has quasi-complete compatibility with the \pandas API\footnote{This assumption is based on a single-machine scenario. It is important to note that while GPUs can provide substantial performance advantages, comparing GPUs and CPUs requires careful consideration of factors like costs (e.g., pay-per-use fees in a cloud environment) and the growing availability of high-core-count CPUs.}.
However, \cudf lacks of an optimizer; thus, libraries like \polars or \sparksql might be faster when running the entire pipeline, avoiding unnecessary materializations~\cite{polars_comparison}
%\footnote{\url{https://pola-rs.github.io/polars/user-guide/misc/alternatives/}}.

% -------------------------------------------------- %

\subsubsection*{Individual stage or complete pipeline?}

If a user is only interested in an individual stage, i.e., input/output (I/O), explorative data analysis (EDA), data transformation (DT), or data cleaning (DC),
\cudf and \polars stand out as the fastest libraries for reading and writing operation respectively.
%\datatable stands out as the fastest library for I/O operations, while for other tasks it might not be that mature.
%Further, it shows the least compatibility with \pandas, requiring to implement many preparators manually.
\polars emerges as the best performer for EDA, while \sparksql represents a consistent solution for DT. 
 Moreover, for DC, \vaex records the best performance among all CPU-only libraries when running the entire pipeline on larger datasets.
% while for DT and DC \sparksql and \vaex show the best results.

% -------------------------------------------------- %

\subsubsection*{What is the size of the datasets?}

For \textbf{small datasets} (less than 500k rows), all of the libraries show similar average runtimes.
Even if \polars is often recommended for such a scenario, and indeed registers the best overall performance if memory resources are not limited, \pandas might still be the most reasonable choice, as their performance does not differ substantially---at least, not enough to replace entire parts of the pipeline~\cite{pinner2023dataframes@medium}.
% In general, if memory resources are not limited, \polars has the best overall performance.

In the case of \textbf{datasets of medium size} (from 2 to 19 million rows) primarily composed of numeric columns (e.g., 5\% \taxi sample), \vaex offers a robust solution if the pipeline involves many column-wise operations and \pandas compatibility is not needed.
On the other hand, for datasets with a high proportion of missing values (e.g., \loan, where more than 30\% of values are missing), \modinr and \pyspark (both version) show good performance.
Finally, for datasets with many string columns (e.g., 25\% \patrol), \sparksql appears to be the best choice.

For \textbf{larger datasets} (over 20 million rows), \sparksql stands out as the best option~\cite{tang2023dataframes@medium}.
Furthermore, when the pipeline requires dealing with heavy-duty column operations, particularly in datasets that have a multitude of string columns and a good amount of null values (e.g., \patrol), \vaex seems to be the best move---and there is no need to set up any environment, as required for \pyspark.

% -------------------------------------------------- %

\section{Conclusion}
\label{sec:conclusion}

We presented a comprehensive experimental comparison of the most used dataframe libraries to support practitioners in the selection of the most suitable solution to carry out their data preparation tasks on a single machine.
To guarantee comparable results and increase usability, we developed \bento, a general framework for assessing the performance of dataframe libraries on four major data preparation stages: input/output (I/O), exploratory data analysis (EDA), data transformation (DT), and data cleaning (DC).

We exploited \bento to perform a thorough comparative analysis of the libraries on four heterogeneous datasets (varying in size, complexity, and features) previously adopted in literature and publicly available on Kaggle.
In particular, we relied on  three of the most voted Kaggle notebooks per dataset to assess the performance of the libraries on real-world data preparation pipelines validated by a large community.
Further, we also evaluate their performance on the popular TPC-H benchmark to support our findings, including \duckdb to assess performance in comparison with an \sql-based system.
%adding {\color{green}\duckdb as reference of \sql based systems.}

%The results of our experiments show that there is no clear winner among the libraries, whose performance is strongly dependent on the size and features of the dataset at hand, on the specifications of the underlying machine, and on the kind of operations to carry out.

%This lack of a "silver bullet" solution underscores the importance of considering multiple factors when choosing a dataframe library for a specific task.
%Our findings show that one size does not fit all in dataframe libraries when it comes to data preparation.
%Yet, we tried to distill takeaways to be used by data practitioners as starting points for the selection of the right library for the data preparation task at hand.
Our experimental evaluation shows that one size does not fit all when it comes to dataframe libraries for data preparation. However, we have distilled key takeaways that data practitioners can use as starting points for selecting the appropriate library for their task at hand.

% {\color{green}
% The inconclusiveness of our results in terms of identifying a single best library actually reflects the reality of data preparation tasks – they are diverse, context-dependent, and often require a tailored approach.
% }
%Thus, we provided the readers with useful takeaways to support them in the selection of the right library for the data preparation task at hand.

As future work, we plan to compare the libraries in a distributed environment and study the possible adoption of machine learning to suggest optimal library combinations based on datasets, tasks, hardware, and previous executions.
Further, we plan to explore how the libraries perform in combination with tools that optimize and rewrite entire dataframe-based data preparation pipelines, such as  \textsf{Dataprep}~\cite{DBLP:conf/sigmod/PengWLBYXCRW21} and \textsf{Dias}~\cite{DBLP:journals/pacmmod/BaziotisKM24}.

% -------------------------------------------------- %

\section*{Acknowledgements}
We extend our appreciation to the master's students of the 2021-2022 Big Data Management and Governance course at the University of Modena and Reggio Emilia for their contributions in testing the libraries.
We warmly thank Leonardo S.p.A., co-funding the PhD program of Angelo Mozzillo.
Further, this work was partially supported by the ``Enzo Ferrari'' Engineering Department (University of Modena and Reggio Emilia) within the projects FARD-2022 and FARD-2023, and by MUR within the project ``Discount Quality for Responsible Data Science: Human-in-the-Loop for Quality Data'' (code 202248FWFS).
% This work was partially supported by the Department of Engineering ``Enzo Ferrari'' within the FARD-2022 (FAR2022DIP\_DIEF - Prot. n. 2985 del 27/07/2022) and FARD-2023 (FAR\_DIP\_2023\_DIEF-SIMONINI) projects.

\bibliographystyle{ACM-Reference-Format}
\bibliography{camera_ready}

%% Please note that all the content must fit within the page limits, including any appendices.
%\appendix
% ...

\end{document}